\documentclass[12pt]{article}
\usepackage{epsfig,multicol}
\usepackage[english]{babel}
\usepackage{graphicx}
\usepackage{rotating}
\usepackage{amsmath}
\usepackage{amssymb}
\usepackage{flafter}

\def\bsg{$b\rightarrow s\gamma$}

\def\Journal#1#2#3#4{{#1} {\bf #2}, #3 (#4)}

\def\NP{Nucl.~Phys.}
\def\PLB{{Phys.~Lett.}~B}

\def\PRL{Phys.~Rev.~Lett.}

\def\PRD{{Phys.~Rev.}~D}

\def\ZPC{{Z.~Phys.}~C}
\def\PTP{{Prog.~Theor.~Phys.}}

\begin{document}

\begin{titlepage}

\begin{flushright}
SISSA 46/2004/EP 
\end{flushright}

\begin{center}

\vspace*{2cm}

{\LARGE A Statistical Analysis of Supersymmetric \\[0.7cm] 

Dark Matter in the MSSM after WMAP}\\

\vspace*{2.5cm}

{\bf\Large S.~Profumo and C.~E.~Yaguna}\\[0.9cm]

         {\em Scuola Internazionale Superiore di Studi Avanzati,
	Via Beirut 2-4, I-34014 Trieste, Italy
and Istituto Nazionale di Fisica Nucleare, Sezione di Trieste, 
I-34014 Trieste, Italy}\\[0.7cm] 

	{\em E-mail:} {\tt profumo@sissa.it, yaguna@sissa.it} 

\vspace*{0.8cm}

\begin{abstract}\noindent
We study supersymmetric dark matter in the general flavor diagonal MSSM by means of an extensive random scan of its parameter space. We find that, in contrast with the standard mSUGRA lore, the large majority of viable models features either a higgsino or a wino-like lightest neutralino, and yields a relic abundance well below the WMAP bound. Among the models with neutralino relic density within the WMAP range, higgsino-like neutralinos are still dominant, though a sizeable fraction of binos is also present. In this latter case, relic density suppression mechanisms are shown to be essential in order to obtain the correct neutralino abundance. We then carry out a statistical analysis and a general discussion of neutralino dark matter direct detection and of indirect neutralino detection at neutrino telescopes and at antimatter search experiments. We point out that current data exclude only a marginal portion of the viable parameter space, and that models whose thermal relic abundance lies in the WMAP range will be significantly probed only at future direct detection experiments. Finally, we emphasize the importance of relic density enhancement mechanisms for indirect detection perspectives, in particular at future antimatter search experiments.
\end{abstract}  
\end{center}
\vspace{1cm}
{\footnotesize PACS numbers: 12.60.Jv, 14.80.Ly, 95.35.+d}

\end{titlepage}


\tableofcontents

\newpage

\section{Introduction}

One of the hints of new physics beyond the standard model of elementary particles (SM) comes from the observation that most of the matter contained in the Universe is non-baryonic. It is therefore a desirable feature of any theory whose purpose is to extend the SM to provide an explanation of this unknown form of matter, commonly dubbed as {\em Dark Matter} (DM).

The minimal supersymmetric extension of the standard model (MSSM) assumes the conservation of a discrete symmetry, called $R$ parity \cite{Fayet:1977yc}, in order to prevent baryon and lepton number violating interactions, which would lead, for instance, to fast proton decay \cite{Weinberg:1981wj,Sakai:1981pk}. The conservation of $R$ parity has the nice feature of rendering the lightest supersymmetric particle (LSP) stable; provided the interactions of the LSP are sufficiently weak, this yields an ideal candidate for dark matter. In particular, it has been shown that the LSP must be electrically and color neutral \cite{coldm}. Among the large plethora of yet unobserved supersymmetric particles of the MSSM, there is a unique candidate for dark matter, the {\em lightest neutralino}.

The recent analysis of the WMAP results on the cosmic microwave background temperature anisotropies, combined with other observational astrophysical data, offered a stringent prediction on the cold dark matter abundance ($\Omega_{\rm CDM}$) within the standard $\Lambda$CDM cosmological model \cite{Spergel:2003cb}\footnote{See \cite{Edsjo:2003us,Ellis:2003cw,Baer:2003yh,Chattopadhyay:2003xi} for recent studies on supersymmetric dark matter in the light of WMAP results within particular SUSY scenarios.}. The greatly improved knowledge of $\Omega_{\rm CDM}$, supplemented with accurate and complementary numerical packages for relic density and detection rates computations in the MSSM \cite{Belanger:2002nx,Gondolo:2002tz}, motivate a further assessment of what is known about neutralino dark matter in the MSSM.

The generic structure of the soft-breaking lagrangian, which dictates most of the phenomenological features of the MSSM, is, however, still largely unknown \cite{Chung:2003fi}. Most studies devoted to supersymmetric dark matter assume either some low-energy relation between the values of the soft-breaking masses \cite{Bergstrom:1995cz,Baltz:2002ei,Bednyakov:2002dz}, or some underlying high-energy principle which organizes the soft terms at the grand unification scale or above \cite{Baer:2000gf,Profumo:2003sx}. The main motivation for this kind of assumptions is that {\em the number of free parameters in the general MSSM is huge} \cite{Chung:2003fi}, and it is practically impossible to draw quantitative predictions out of such an unmanageable parameter space.

The purpose of the present study is to make statistical statements about the implications of the stringent WMAP bounds on the mass and composition of the lightest neutralino and on dark matter detection perspectives, in the {\em most general realization of the MSSM} compatible with all phenomenological constraints. No {\em a priori} relations will be assumed among the soft breaking masses and parameters appearing in the lagrangian of the theory. 

The main tool we make use of is a very large random scan of the MSSM parameter space, from which we extract a sufficiently wide set of {\em viable models}, i.e. models which are consistent not only with all phenomenological constraints, but which also produce a neutralino relic abundance within the {\em upper} bound of the WMAP CDM allowed range.

Though the relevant number of parameters we use in our scan is rather large (20 overall, see Sec.~\ref{sec:ps}), supersymmetric models which feature the correct relic density within the WMAP range mainly fall into three categories (Sec.~\ref{sec:nature}): 
\begin{itemize}
\item[-] {\em higgsino}-like LSP with a mass around 1 TeV.
\item[-] {\em wino}-like LSP with a mass around 1.6 TeV.
\item[-] {\em bino}-like LSP, with a wide mass range (up to the TeV region).
 \end{itemize}
Since in the case of a bino-like lightest neutralino coannihilations or resonances are, as a matter of fact, unavoidable, we consider in great detail the various coannihilating partners and their relative effectiveness, as well as the general features of the resonant heavy Higgs boson $s$-channel exchanges (Sec.~\ref{sec:mech}). 

Finally, we study the dark matter detection perspectives in a number of search channels: spin dependent and spin-independent direct detection, indirect detection at neutrino telescopes, and at antiprotons and positrons search experiments (Sec.~\ref{sec:det}). The aim of our analysis is to transparently compare different search strategies. To this extent, we make use of {\em self consistent halo models}, motivated by available numerical and observational data, and we present our results in the form of {\em visibility ratios}, i.e. signal-to-sensitivity ratios. 

We concentrate our analysis on two cases: models in which the neutralino thermal relic density is also consistent with the lower WMAP bound, and models which only fulfill the upper bound, and which, in general, have an excessively low relic density in the standard cosmological scenario. In this latter case, we assume the existence of {\em relic density enhancement mechanisms} which could affect the standard thermal relic abundance computations, and render models with large annihilation rates (and therefore low relic densities) compatible even with the lower WMAP bound. 

In this respect, a wealth of scenarios have been proposed, such as {\em non-thermal production} of neutralinos \cite{non-therm}, cosmological enhancements due to {\em quintessential effects} \cite{quint,Profumo:2003hq}, to {\em anisotropic cosmologies} producing an effective shear energy density \cite{shear, Profumo:2004ex}, or to {\em scalar-tensor theories} \cite{Catena:2004ba}. Under the assumption of the existence of such enhancement mechanisms, no rescaling procedure is in order, for models underproducing dark matter in the standard cosmological scenario of thermal production. Indirect detection rates, which essentially depend on the same annihilation rates determining the neutralino relic abundance, will then be considerably larger than for model whose relic density falls within the WMAP range. 

We point out that current data on dark matter detection {\em do not significantly constrain the viable supersymmetric parameter space}, at least in a statistical sense. Regarding future perspectives, spin-independent direct detection is the most promising search strategy for models whose neutralino relic abundance lies in the WMAP range. We also find that neutrino telescopes have a limited parameter space reach, and can only probe models with relatively low masses. We however point out the correlations and complementarity between neutrino telescopes searches and direct dark matter detection. Finally, we emphasize that models with large annihilation rates, assuming some relic density enhancement mechanism, give spectacular signals at antimatter searches, and are even already constrained by available data.

\section{MSSM Parameter Space Scan}\label{sec:ps}

\subsection{Motivation}
Supersymmetry, if it exists, must be spontaneously broken. The precise mechanism of supersymmetry breaking is, however, not known. From a practical point of view this difficulty is overcome by introducing into the MSSM effective Lagrangian extra soft terms which break supersymmetry explicitly. The collection of such terms is known as the {\em soft-breaking Lagrangian} (see \cite{Chung:2003fi} for a recent review), and reads 
\begin{eqnarray}
\label{softbreaking}
-{\cal{L}}_{soft}&=&(m_{\tilde Q}^2)_i^j {\tilde q}_{L}^{\dagger i}
{\tilde q}_{Lj}
+(m_{\tilde u}^2)^i_j {\tilde u}_{Ri}^* {\tilde u}_{R}^j
+(m_{\tilde d}^2)^i_j {\tilde d}_{Ri}^* {\tilde d}_{R}^j
\nonumber \\
& &+(m_{\tilde L}^2)_i^j {\tilde l}_{L}^{\dagger i}{\tilde l}_{Lj}
+(m_{\tilde e}^2)^i_j {\tilde e}_{Ri}^* {\tilde e}_{R}^j
\nonumber \\
& &+{\tilde m}^2_{h1}h_1^{\dagger} h_1
+{\tilde m}^2_{h2}h_2^{\dagger} h_2
+(B \mu h_1 h_2 + h.c.)
\nonumber \\
& &+ ( A_d^{ij} Y_d^{ij} h_1 {\tilde d}_{Ri}^*{\tilde q}_{Lj}
+A_u^{ij} Y_u^{ij} h_2 {\tilde u}_{Ri}^*{\tilde q}_{Lj}
+A_l^{ij} Y_l^{ij}h_1 {\tilde e}_{Ri}^*{\tilde l}_{Lj}
\nonumber \\
& & +\frac{1}{2}M_1 {\tilde B}_L^0 {\tilde B}_L^0
+\frac{1}{2}M_2 {\tilde W}_L^a {\tilde W}_L^a
+\frac{1}{2}M_3 {\tilde G}^a {\tilde G}^a +h.c.).
\end{eqnarray}
which respectively corresponds, line by line, to squark masses, slepton masses, Higgs masses, trilinear couplings, and gaugino masses. 
The most general ${\cal{L}}_{soft}$ introduces a huge number of free parameters (more than 100), which makes any phenomenological study based on a blind approach to the MSSM parameter space very problematic. The usual way of studying supersymmetric effects is to assume {\em specific frameworks} in which all soft-parameters depend on a few inputs given at a certain high energy scale. Among them, minimal supergravity (mSUGRA) models \cite{msugra} have received particular attention. mSUGRA models are defined in terms of four continuous parameters and one sign:
\begin{equation}\label{msugra}
\tan\beta,\quad m_0 ,\quad a_0,\quad m_{1/2},\quad \mathrm{sign}(\mu)\,,
\end{equation} 
which determine the whole set of soft breaking terms at the unification scale $M_{GUT}$ through the relations:
\begin{equation}
M_i=m_{1/2}, \quad(m^2_{\tilde f})_{ij}= m_0^2 \delta_{ij},\quad\tilde m_{h1}^2=\tilde m_{h2}^2=m_0^2,\quad A_f=a_0\,.
\end{equation}
Thus, at $M_{GUT}$ all scalars have a common mass $m_0$, and all gauginos feature the same soft mass $m_{1/2}$.

The allowed parameter space of mSUGRA models, compatible with all phenomenological and cosmological constraints have been determined in a number of works (among recent studies see e.g.~\cite{Edsjo:2003us,Baer:2003wx}). It turns out that the most important constraint comes from the requirement that the neutralino relic density does not exceed the relic density of cold dark matter. 

In mSUGRA, the lightest neutralino is usually a bino-like neutralino, which has a small annihilation cross section, and therefore tends to produce a large relic abundance. Indeed, apart from a small region at low neutralino mass, only three regions fulfill the WMAP upper bound on the relic density:

\begin{itemize} 
\item The stau coannihilation strip: Along this region the stau is almost degenerate with the lightest neutralino, and coannihilation processes help suppress the relic density.
\item The Funnel region: In this case there is a resonant enhancement of the bino-bino annihilation cross section through s-channel heavy Higgs bosons exchange. A necessary condition required for such a resonant enhancement is a large value of $\tan\beta$, needed to fulfill the relation $2 m_\chi\approx m_A$.
\item The focus point region: These are narrow regions with very large values of $m_0$ yielding, through electroweak symmetry breaking conditions, a low value of the $\mu$ parameter, and implying a lightest neutralino with a non-negligible, or even a large higgsino fraction. 
\end{itemize}

A legitimate question one can ask at this point is, therefore, whether or not mSUGRA is a {\em benchmark scenario} for SUSY dark matter with respect to the bulk of the general MSSM parameter space, or if it is a somehow {\em theoretically biased framework}. In this paper we try to give an answer to this issue too, taking a different approach to the study of supersymmetric models. Instead of choosing a particular model of SUSY breaking, we carry out a statistical analysis of a huge number of models with randomly generated soft parameters. We do not make any simplifying hypothesis for the low energy parameters, and we therefore do not assume any scalar universality or gaugino unification relations. In this way we wish to get a feeling of the typical predictions of SUSY models, and to compare them against mSUGRA or any other specific scenario.

\subsection{The Scan}

Many of the parameters of ${\cal{L}}_{soft}$ are severely constrained because they would imply FCNC or CP violating effects at a rate which is already ruled out by the experiments. For simplicity, we will assume, as usually done in the literature, that all soft parameters are real, so that supersymmetry breaking does not introduce new sources of CP violation. To suppress potentially dangerous FCNC, we will set to zero all off-diagonal elements in the sfermions masses, and assume that the first and second generation of squarks are degenerate (for a recent discussion see \cite{Baer:2004xx}). Squarks masses therefore have the form
\begin{equation}
m^2_{\tilde Q}=\left(\begin{array}{ccc} m_q^2 & &\\ & m_q^2 & \\ & & m_{Q3}^2\end{array}\right),\;
m^2_{\tilde u}=\left(\begin{array}{ccc} m_q^2 & &\\ & m_q^2 & \\ & & m_{u3}^2\end{array}\right),\;
m^2_{\tilde d}=\left(\begin{array}{ccc} m_q^2 & &\\ & m_q^2 & \\ & & m_{d3}^2\end{array}\right),
\end{equation}
with $m_q^2$, $m_{Q3}^2$, $m_{u3}^2$ and $m_{d3}^2$ arbitrary numbers. Slepton masses, on the other hand, contain three independent entries each.

The trilinear couplings $A_t$, $A_b$, $A_\tau$ and $A_\mu$ are allowed to have both signs. All other trilinear couplings are neglected\footnote{We include $A_\mu$ for its relevance in the computation of $(g-2)_\mu$.}.
Gaugino masses are independent of one another, and negative values for $M_2$ are also considered. The remaining low energy parameters we take into account are $\tan\beta$, $\mu$ and $m_A$, the mass of the pseudo-scalar Higgs boson. 

\begin{table}[!h]
\begin{center}
\begin{tabular}{|l|c|c|r|}\hline
Name & \# of parameters & Symbol & Range~~~~~~~~~\\
\hline
 Gluino and Bino Masses& 2 & $M_1$,$M_3$ & ($50$ GeV, $5$ TeV)\\
Wino Mass& 1 & $M_2$ & $\pm$ ($50$ GeV, $5$ TeV)\\
Left-handed slepton masses & 3 & $m_{\tilde l_i}$& ($50$ GeV, $5$ TeV)\\
Right-handed slepton masses & 3 & $m_{\tilde e_i}$& ($50$ GeV, $5$ TeV)\\
1st and 2nd family squark masses & 1 & $m_{q}$ & ($50$ GeV, $5$ TeV)\\
3rd family squark masses & 3 & $m_{\tilde u_3}$,$m_{\tilde d_3}$,$m_{\tilde Q_3}$&($50$ GeV, $5$ TeV)\\
Third family trilinear couplings & 3 & $A_t,A_b,A_\tau$& ($-5$ TeV, $5$ TeV)\\
Pseudo-scalar Higgs boson mass & 1 & $m_A$ &($50$ GeV, $5$ TeV)\\
Muon trilinear coupling & 1 & $A_\mu$ & ($-5$ TeV, $5$ TeV)\\
$\mu$ & 1 & $\mu$ & $\pm$($50$ GeV, $5$ TeV)\\
$\tan \beta$ & 1 & $\tan \beta$ & ($2$, $50$)\\
\hline
Total & 20 & &\\
\hline  
\end{tabular}
\end{center}
\caption{List of the MSSM parameters taken into account in our scan, and the range on which they are allowed to vary.}\label{table:param}
\end{table}
 
We scan the resulting  20-dimensional parameter space using a {\em uniform probability distribution}. $\tan\beta$ takes values between $2$ and $50$, whereas all mass parameters are generated in the interval $(50\, \mbox{GeV},5\, \mbox{TeV})$ possibly with both signs, according to Table~\ref{table:param}.
From this set of low energy parameters, one can determine the mass spectra and mixing matrices of the superparticles. 

All the results we present in this paper are obtained from this kind of procedure. Had we chosen to scan over a smaller number of parameters (assuming additional relations), or to drastically change the range on which they vary, the results might have been different. We claim however that this scan covers what can be considered a natural parameter space range, and that it is largely free of theoretical prejudices.

\subsection{Phenomenological Constraints}
We apply the following phenomenological constraints:
\begin{itemize}
\item {\em The spectrum:} The presence of non-zero trilinear couplings could give rise to tachyonic sfermions, so the first consistency requirement we ask is to exclude all such unphysical models.
\item {\em Neutralino LSP:} In the MSSM the lightest supersymmetric particle is stable and therefore must be an electrically neutral and not strongly interacting particle \cite{coldm}. Since the sneutrino, in the MSSM, has been shown not to be a suitable dark matter candidate \cite{Falk:1994es}, the only possibility we are left with is the lightest neutralino. We therefore also require the LSP to be the lightest neutralino.

\item $b\rightarrow s\gamma$: In SUSY models with minimal flavor violation the decay $b\rightarrow s\gamma$ proceeds through the $\tilde t \tilde W$ and $tH^+$ loops, in addition to the SM contribution from the $tW$ loop. The branching fraction $BF(b\rightarrow s\gamma)$ has been measured by the BELLE, ALEPH, and CLEO collaborations. A weighted averaging of these measurements of $B\rightarrow X_s\gamma$ decays at CLEO and BELLE lead to bounds on the branching ratio $b\rightarrow s\gamma$. We will require
\begin{equation}
2\times 10^{-4}\leq BR(b\rightarrow s\gamma)\leq 4.6\times 10^{-4}.
\end{equation}
The calculation of $BR(b\rightarrow s\gamma)$ is carried out with the latest release of the \texttt{micrOMEGAs} package \cite{Belanger:2004yn}, which includes an improved NLO and the charged Higgs contributions as well as a beyond leading-order treatment of large $\tan\beta$ effects.
\item {\em Direct Accelerator Searches}: We also take into account the limits derived from the unsuccessful searches for supersymmetric particles and the Higgs boson. In Table \ref{table:limit} we detail on the mass limits we impose on each SUSY particle.
\begin{table}[!h]
\begin{center}
\begin{tabular}{lrrl}
Name & Mass & Limit &(GeV) \\
\hline
Charginos & &103.5 & \\
sneutrinos & &43.0 & \\
charged sleptons & & 95.0& \\
sbottoms & & 91.0& \\
stops & & 86.4 & \\
other squarks & & 100.0& \\
gluino & & 195.0& \\
Higgs boson & & 114.0& \\
Pseudoscalar Higgs boson & & 85.0& \\ 
\hline
\end{tabular}
\end{center}
\caption{Mass limits}\label{table:limit}
\end{table}

\item {\em Relic density}: Finally, we require that the neutralino relic density lies below the WMAP $2-\sigma$ upper bound \cite{Spergel:2003cb},
\begin{equation}
\Omega_{DM} h^2 \leq 0.13. 
\end{equation}
The WMAP lower bound ($\Omega_{DM} h^2 \geq 0.09$) is not imposed since it may well be that neutralinos are  only a subdominant dark matter component. 
For the evaluation of the neutralino relic density we again make use of the program \texttt{micrOMEGAs}, which takes into account all tree level annihilation processes as well as all possible coannihilation processes in the MSSM. We also cross-checked our statistical results for a sample of models with the {\tt DarkSUSY} package, and verified that they are overall consistent with each other. 
\end{itemize}
\begin{figure*}[!t]
\begin{center}
\includegraphics[scale=0.7]{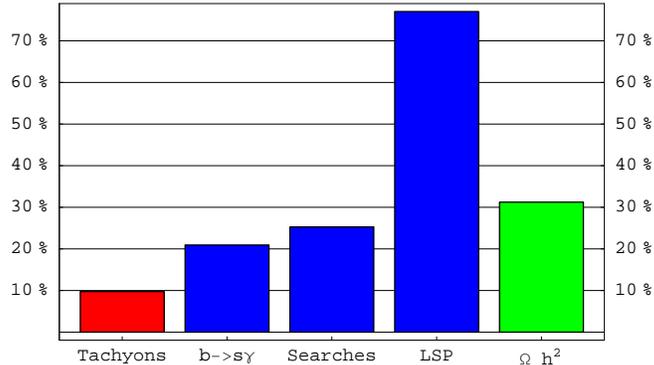}
\end{center}
\caption{\small \em Effectiveness of the different constraints imposed on supersymmetric models.}\label{fig:excl}
\end{figure*}

Although our main focus is on supersymmetric Dark Matter, we will  first use our scan to {\em quantify} the effectiveness of the different constraints mentioned above. The results are illustrated in  Fig.~\ref{fig:excl}. Different colors indicate different samples: Red for the whole sample, Blue for those models {\em surviving} the red constraint, and green for those surviving the blue one. We see that $10\%$ of the models contain a tachyon; of the remaining models, $20\%$ are ruled out by the $b\rightarrow s\gamma$ constraint, $25\%$ are excluded by direct searches and almost $80\%$ contain a  LSP different from a neutralino. Finally, $30\%$ of the models which are not excluded by any of the previous tests give rise to a neutralino relic abundance larger than the WMAP upper bound. The surviving models will be called {\em viable}, and are the only ones we will consider from now on. Our sample consists of $10^5$ viable models.

\subsection{A statistical analysis of \bsg, $m_h$ and $(g-2)_\mu$}

\begin{figure*}[!t]
\begin{center}
\begin{tabular}{ccc}
\includegraphics[scale=0.75]{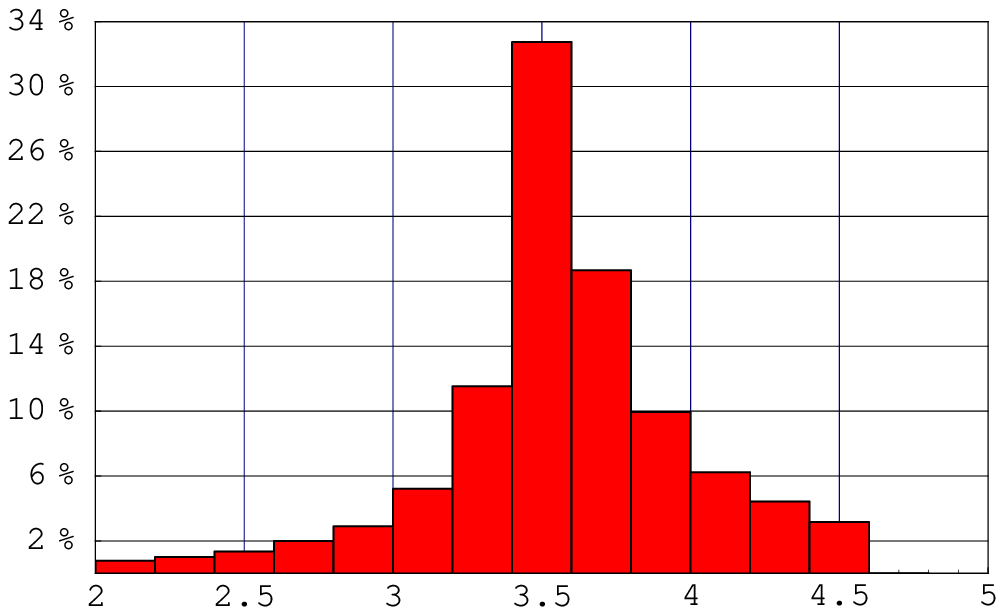} & \hspace{0.25cm} &\includegraphics[scale=0.75]{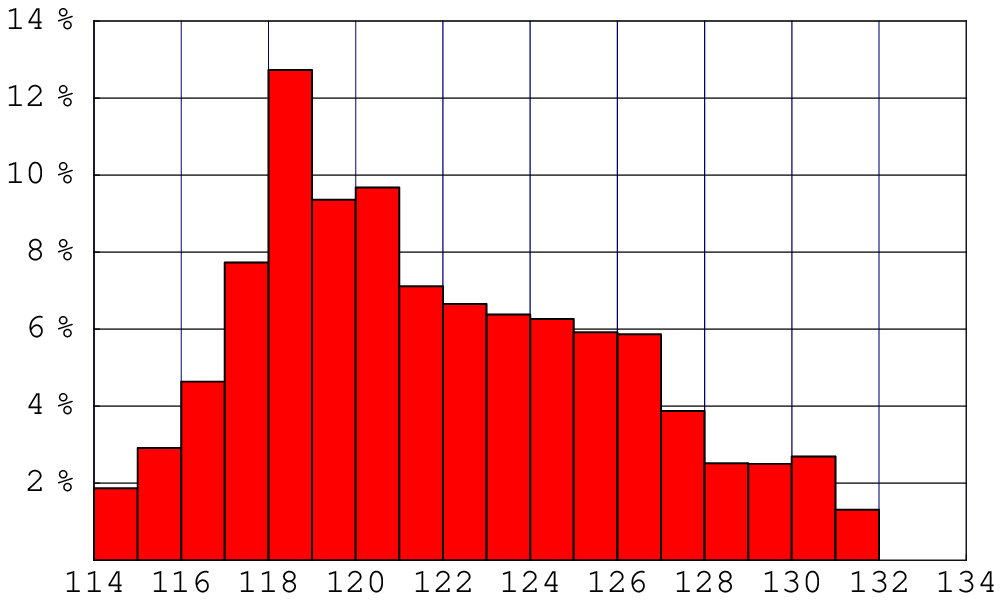}\\
({\em a\ }) && ({\em b\ })\\
\end{tabular}
\end{center}
\caption{\small \em (a): $BF(b\rightarrow s\gamma)$ in units of $10^{-4}$. (b): The Higgs boson mass, in GeV. An upper bound $m_h\leq 132$ emerges from our scan.}\label{fig:bsgmh}
\end{figure*}

We take at this point the opportunity to assess the statistical results concerning three quantities sensitive to supersymmetric effects, i.e. the rare decay \bsg, the mass of the lightest $CP$-even Higgs boson and the muon anomalous magnetic moment $(g-2)_\mu$.

In panel (a) of Fig.~\ref{fig:bsgmh} we show a histogram with the value of $BF(b\rightarrow s\gamma)$ in units of $10^{-4}$. Notice that most models predict a value close to the central value  predicted by the SM ($\approx 3.4$). That is, the supersymmetric contributions to $b\rightarrow s\gamma$ tend to be  very small, typically signaling a heavy SUSY spectrum. On average, the SUSY contributions are found to be {\em positive}, and to increase, therefore, the SM value of $BF(b\rightarrow s\gamma)$.

Among the different limits shown in Tab.~\ref{table:limit} the Higgs Boson mass plays a special r\^ole. Indeed, one of the generic predictions of the MSSM is a light, ``standard-model-like'' Higgs boson. The computation of the Higgs mass is carried out with the {\tt FeynHiggsFast} package, which includes radiative corrections up to the two-loop level. For the top mass we set $m_t=175$ GeV. In Fig.~\ref{fig:bsgmh} (b) we show a histogram of the Higgs boson mass in units of GeV. Notice that there is a smooth peak around $m_h\approx 119$ GeV, and that no models were found with $m_h> 132$ GeV (this limit is very similar to what reported in earlier analysis, see e.g.~\cite{hcorrections}). 

\begin{figure*}[!t]
\begin{center}
\includegraphics[scale=0.8]{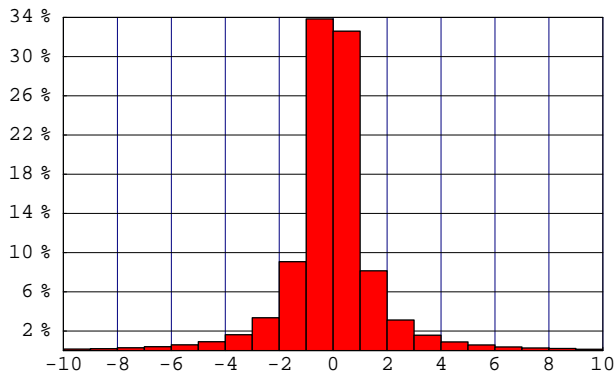}
\end{center}
\caption{\small \em The supersymmetric contribution to the anomalous magnetic moment of the muon ($a_\mu=(g-2)_\mu/2$) in units of $10^{-10}$. }\label{fig:gmuon}
\end{figure*}

Due to a long standing uncertain theoretical and experimental situation, we did not take into account any bound deriving from the anomalous magnetic moment of the muon. Nevertheless, we show in Fig.~\ref{fig:gmuon} a histogram of the supersymmetric contribution to $a_\mu=(g-2)_\mu/2$. The calculation is done with the routine provided with \texttt{micrOMEGAs}. It is clear from the figure that most models predict a small value of $\delta a_\mu$:
\begin{equation}
-2.0\times 10^{-10}<\delta a_\mu<2.0\times 10^{-10}\,.
\end{equation}  
Using the latest experimental and theoretical results, based on data from $e^+e^-\rightarrow hadrons$, the deviation $\delta a_\mu$ was found to be \cite{Baer:2004xx}:
\begin{eqnarray}
\delta a_\mu&=&(27.1\pm9.4)\times 10^{-10}\qquad \text{(Davier {\em  et al.}) \cite{Davier}}\\
\delta a_\mu&=&(31.7\pm9.5)\times 10^{-10}\qquad \text{(Hagiwara {\em  et al.}) \cite{Hagiwara}}
\end{eqnarray}
If, instead, the $\tau$ decay data is used to determine the hadronic vacuum polarization, the deviation is smaller:
\begin{equation}
\delta a_\mu=(12.4\pm 8.3)\times 10^{-10}\qquad \text{(Davier {\em et al.}) \cite{Davier}}\,.
\end{equation}
In a conservative approach, such as that outlined in Ref.~\cite{Martin:2002eu}, we can therefore conclude that {\em the bounds stemming from $a_\mu$ only constrain a marginal portion of the viable  MSSM parameter space}.

\section{The Nature of Neutralino DM after WMAP}\label{sec:nature}

\subsection{Relic Density of Neutralino DM}
\begin{figure*}[!t]
\begin{center}
\includegraphics[scale=0.6]{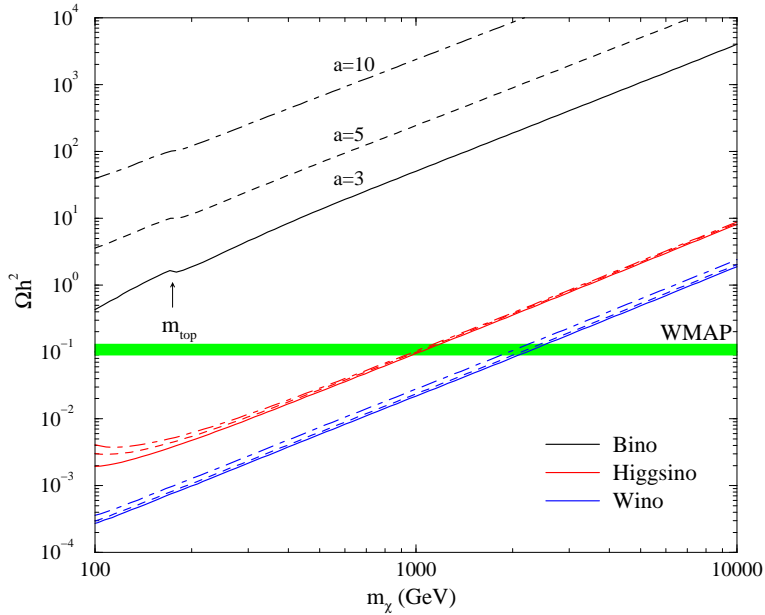}
\end{center}
\caption{\small \em  The neutralino relic density as a function of $m_\chi$ for a  bino-, a wino- and a higgsino-like neutralino. All other superparticles are assumed to have a mass much larger than $m_\chi$: $m_{susy}\approx a\cdot m_{\chi}$, with $a=3,5,10$.}\label{fig:NEUT}
\end{figure*}

The relic density of the lightest neutralino depends critically on two factors: the neutralino {\em mass} $m_\chi$ and the neutralino {\em interactions}. In the MSSM, the lightest neutralino is a linear combination of the gauge eigenstates, which are dubbed bino, wino and higgsino; its interactions are therefore determined  by the relative bino-, wino- and higgsino-content. We show in Fig.~\ref{fig:NEUT} a plot of the relic density as a function of the neutralino mass for a bino-, a higgsino- and a wino-like neutralino. All other superparticles are assumed to have a mass much larger than $m_\chi$ and given by $m_{\rm SUSY}\approx a\cdot m_\chi$ (Notice that doing so, we are neglecting the possibility of coannihilations or resonances through an $s$-channel\footnote{Assuming that the scalar masses lie well above the gaugino masses amounts to resorting to a scenario similar to that of the {\em finely tuned MSSM} \cite{Arkani-Hamed:2004fb}, sometimes also dubbed {\em Split Supersymmetry}, see \cite{Giudice:2004tc}.}). From the plot we see that the dependence with $a$ is particularly relevant for a bino-like neutralino. Some general conclusions can be drawn from this plot:
\begin{itemize}

\item A bino-like neutralino typically produces a relic abundance that is much larger than the WMAP bound. In fact, we will see that  viable models with bino-like neutralinos require the presence of peculiar mechanisms which suppress the relic density.

\item A higgsino-like neutralino must have a mass of the order of $1$ TeV in order to produce a relic abundance consistent with the preferred WMAP range.

\item A wino-like neutralino generates a relic abundance in the WMAP range only provided $m_\chi\sim 2$ TeV. 
\end{itemize} 

\begin{figure*}[!t]
\begin{center}
\includegraphics[scale=1.0]{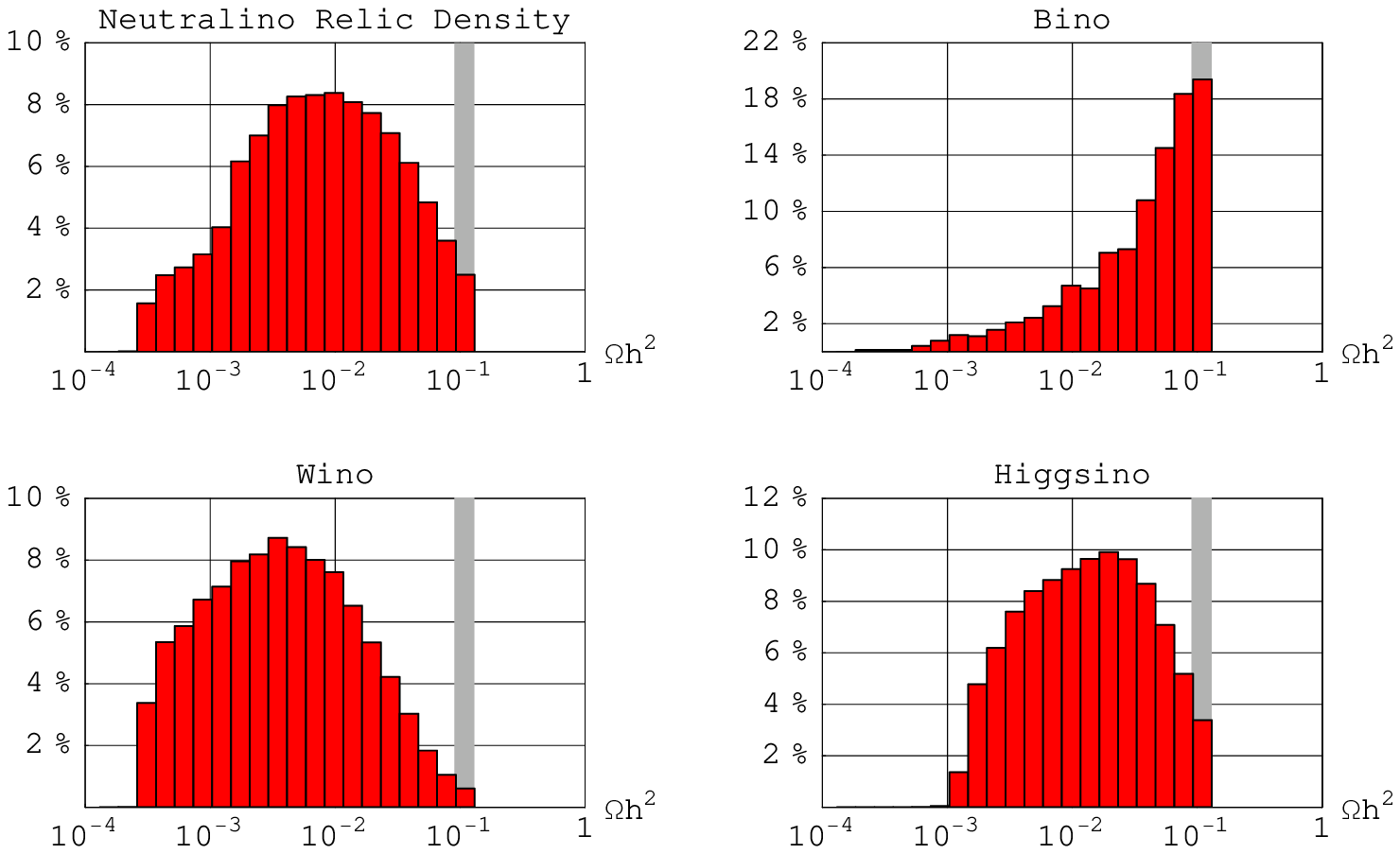}
\end{center}
\caption{\small \em Histograms of the neutralino relic density. The first panel shows the statistical distribution of the neutralino relic density for all models. In the other panels, the relic density of a bino-, a wino- and a higgsino-like neutralino are shown. The gray vertical band corresponds to the WMAP range.}\label{fig:omeall}
\end{figure*}

In Fig.~\ref{fig:omeall} we show a histogram of the total neutralino relic density, and  of the relic density of bino-, wino- and Higgsino-like neutralinos.  A neutralino is said to be bino-like if its bino component is larger than its wino and higgsino components. Wino- and higgsino-like neutralinos are defined analogously. From the first diagram, we learn that most models produce a relic density that is {\em much smaller than the WMAP bound} (the gray vertical band), with a moderate peak around $\Omega_\chi h^2\approx 10^{-2}$. This picture is clearly very different from the usual situation in supergravity inspired frameworks, where the neutralino relic abundance is typically found to be excessively large. The second diagram reveals that bino-like neutralinos typically yield a large relic abundance, and that they contribute significantly to the WMAP range. Models with wino- and higgsino-like  neutralinos, on the other hand, feature a relic density well below the WMAP range, as apparent from the Wino and Higgsino histograms.

\subsection{Neutralino Mass and Composition}

\begin{figure*}[!t]
\begin{center}
\begin{tabular}{ccc}
\includegraphics[scale=0.5]{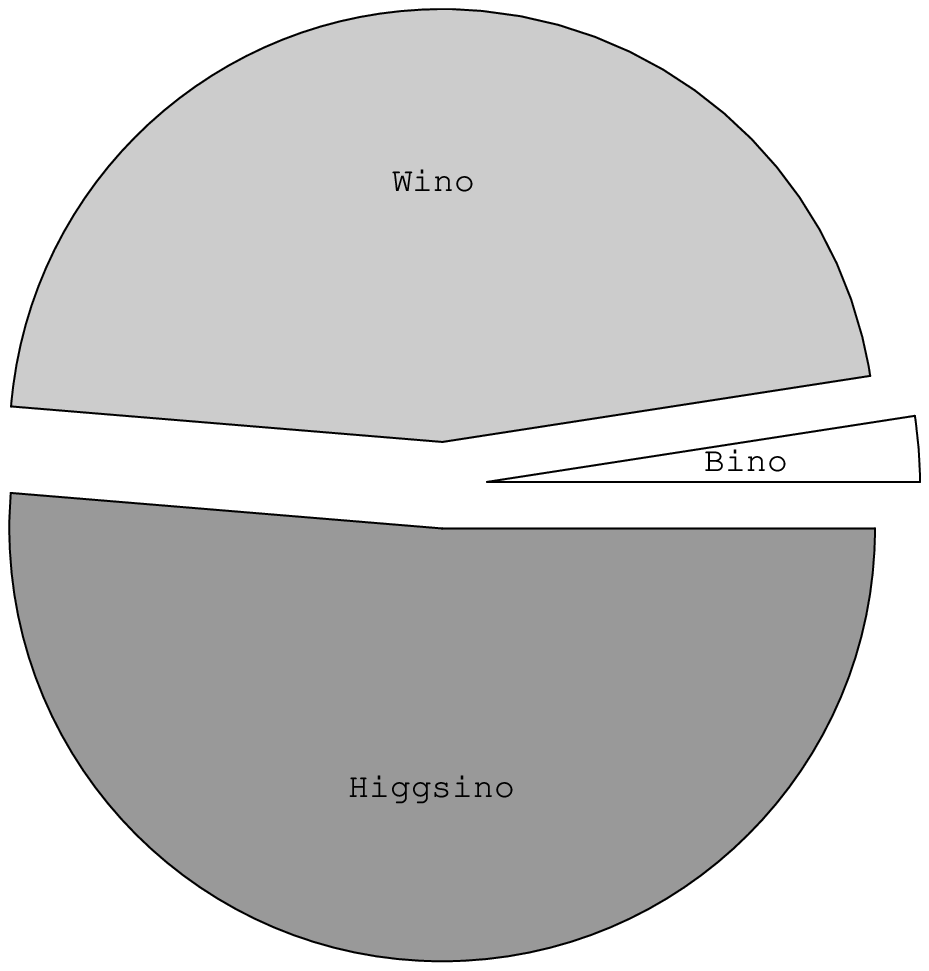} & \hspace{2cm} &\includegraphics[scale=0.5]{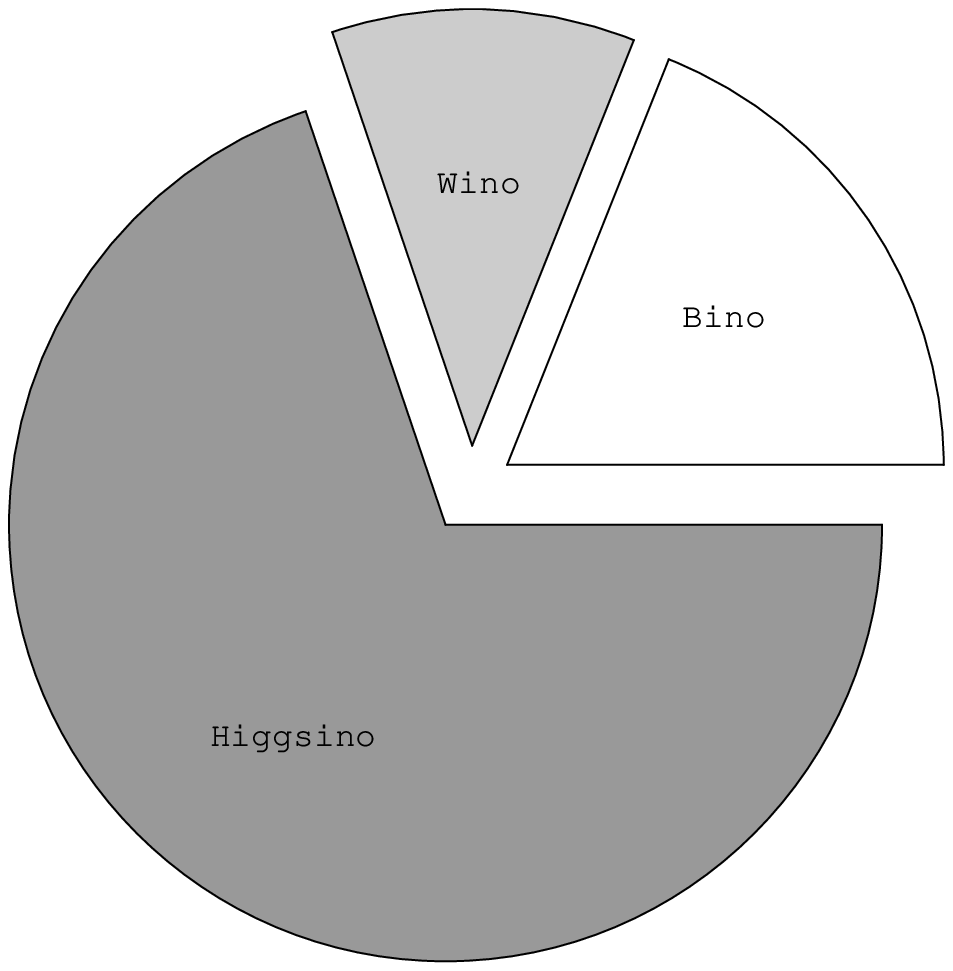}\\
({\em a\ }) && ({\em b\ })\\
\end{tabular}
\end{center}\caption{\small \em Pie diagrams showing the composition of the lightest neutralino. In $(a)$ we include all viable models; in $(b)$, only those with a relic density within the WMAP range.}\label{fig:pie}
\end{figure*}
\begin{figure*}[!t]
\begin{center}
\includegraphics[scale=1.0]{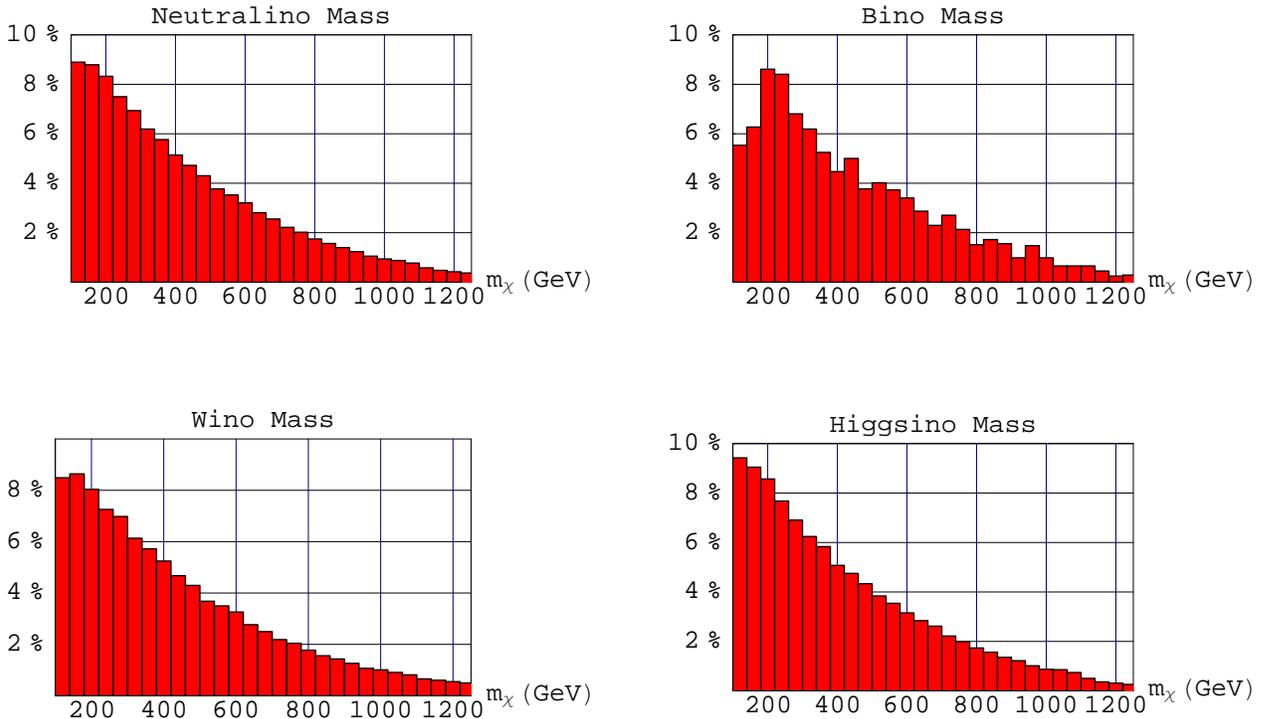}
\end{center}
\caption{\small \em Histograms of the neutralino mass. The first panel shows the mass of the neutralinos for all viable models. In the others, we examined each kind of neutralino separately.}\label{fig:nmasses}
\end{figure*}

\begin{figure*}[!t]
\begin{center}
\includegraphics[scale=1.0]{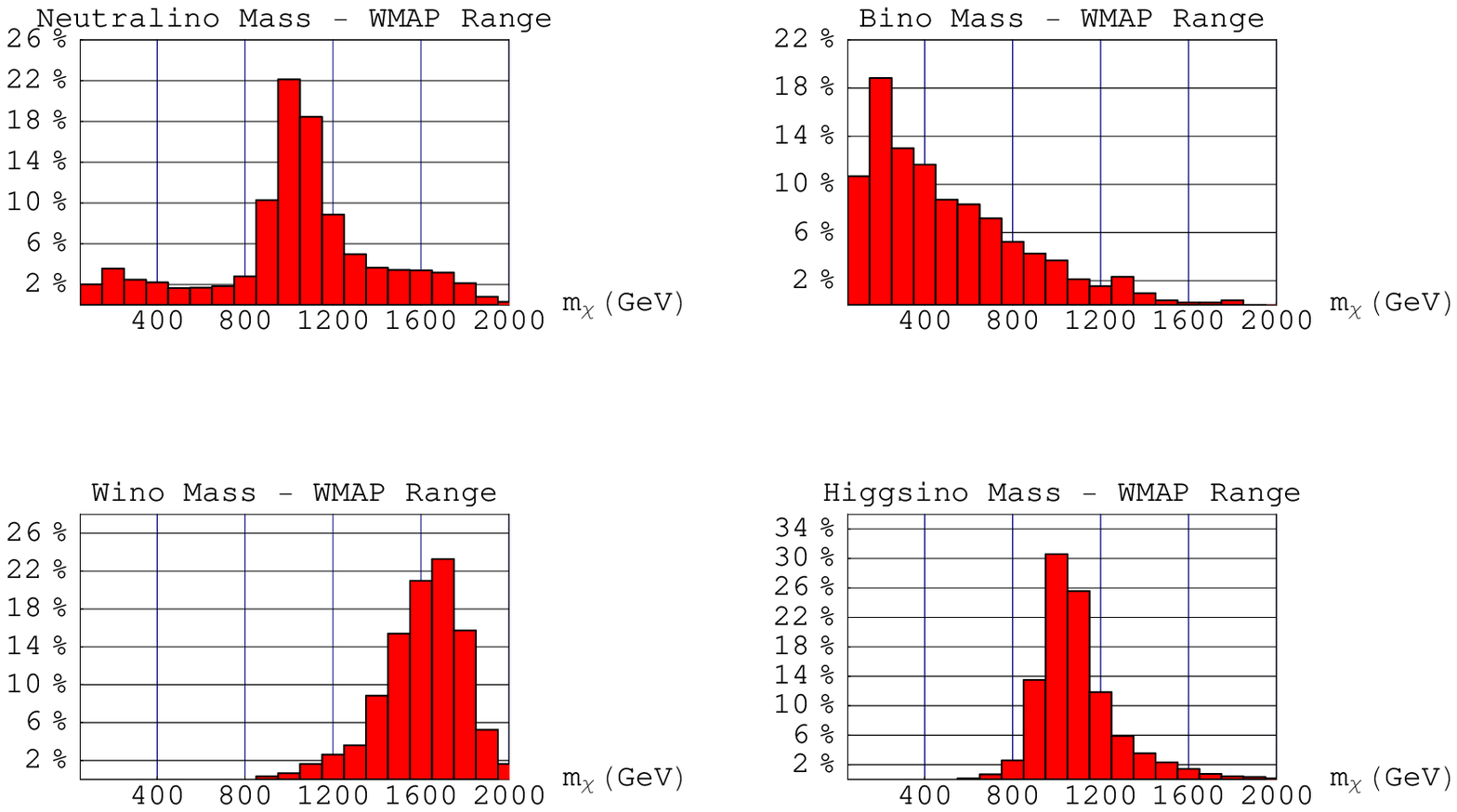}
\end{center}
\caption{\small \em Neutralino masses giving a relic density in the WMAP range.}\label{fig:nmasseswmap}
\end{figure*}

The composition of a typical neutralino markedly depends on whether or not we demand the WMAP lower bound to be fulfilled. This fact is illustrated in Fig.~\ref{fig:pie} with pie diagrams showing the relative number of bino-, wino-, and higgsino-like neutralinos. In $(a)$, where the lower bound was not imposed, the lightest neutralino is predominantly a higgsino or a wino, with only a small fraction of binos. In $(b)$, where the relic density is required to lie {\em within the WMAP range},  the lightest neutralino is mostly a higgsino, with a sizeable fraction of binos and a moderate contribution from winos.

In Fig.~\ref{fig:nmasses} we show  histograms of the lightest neutralino mass for all viable models.  As expected from statistical reasons, the probability of having a neutralino LSP with a given mass $m$ decreases with $m$. Notice that the mass probability distribution $P(m_\chi)$ of a wino (and of a higgsino) follows the statistically expected functional form, dictated by the requirement that $M_2$ (respectively $\mu$) is the lightest among $n_P\simeq 15$ mass parameters which range in the $[m_{\rm UP},m_{\rm DOWN}]$ interval, i.e. the following power-law:
\begin{equation}
P(m_\chi)\propto\left(m_{\rm UP}-m_\chi\right)^{(n_P-1)}.
\end{equation} 
On the other hand, the distribution we find for the bino masses reflects the one obtained under the requirement of having at least one coannihilating partner, whose mass lies between $m_\chi$ and $(1+\delta)m_\chi$, i.e.
\begin{equation}
P^C(m_\chi)\propto \left(m_{\rm UP}-m_\chi\right)^{(n_P-1)}\cdot \left(1-\left(\frac{m_{\rm UP}-(1+\delta)m_\chi}{m_{\rm UP}-m_\chi}\right)^{(n_P-1)}\right)
\end{equation}  
which, in the limit of small $\delta$, features a maximum at
\begin{equation}
m_\chi=m_{\rm UP}/(n_P-1)\approx 350\, {\rm GeV},
\end{equation}
in reasonable agreement with what we statistically find in the case of binos\footnote{The location of the maximum in the histogram is at slightly lower masses due to the presence of non-coannihilating light binos.}. This is clearly a first strong indication of the (statistical) relevance of coannihilations for bino-like neutralinos; although coannihilations may appear as a finely-tuned mechanism of relic density suppression, cosmology seems to point, instead, to their ``naturalness''. We will carry out a more detailed analysis of this point in Sec.~\ref{sec:mech}.

In Fig.~\ref{fig:nmasseswmap} we show histograms of the neutralino mass for models with a relic density in the WMAP range. We see that most models have a neutralino mass in the range $800$ GeV$<m_\chi<1200$ GeV. Binos are usually light ($m_\chi<1 $ TeV) whereas winos and higgsino masses cluster around $m_\chi\approx 1.6 $ TeV and $m_\chi\approx 1 $ TeV. Since the LHC reach for higgsino-like or wino-like neutralinos will likely be well below an LSP mass of 1 TeV, we point out that {\em SUSY models with a relic abundance lying within the WMAP range and featuring a wino or higgsino-like lightest neutralino, will not be detectable at the LHC}.

\section{Relic Density Suppression Mechanisms}\label{sec:mech}

In the previous sections we pointed out the well known fact that the neutralino relic density critically depends on the composition of the LSP in terms of its bino, wino and higgsino components. The annihilation cross section of a bino is by far smaller than those of a wino or a higgsino: first, the couplings involving a bino are smaller than those, for instance, of a wino, since $g_1<g_2$; furthermore, the number of final states in which a bino can annihilate is much smaller than that of winos and higgsinos, and the latter can directly annihilate, without the need of an intermediate supersymmetric partner, into, for example, a couple of gauge bosons, whereas binos cannot. This translates into a critical dependence of bino annihilation cross section on the supersymmetric spectrum, which is, on the other hand, absent in the case of winos and higgsinos. Last but not least, the mass matrix structure of charginos and neutralinos implies that a wino-like LSP has a quasi degenerate lightest chargino, yielding a large coannihilation contribution; for the same reason, higgsinos are quasi degenerate both with the lightest chargino and with the next-to-lightest neutralino, again with large corresponding coannihilation effects. In this respect, the features of Fig.~\ref{fig:NEUT} come not as a surprise: the relic density of higgsinos and winos has a weak dependence on the SUSY particle spectrum, and is orders of magnitude suppressed with respect to that of binos.

The bottom line is, on the one hand, that the mass spectrum of higgsinos and winos whose relic density falls within the WMAP range clusters around definite values of the neutralino mass, the only relevant parameter. On the other hand, we will hereafter show that, in most cases, binos are compatible with the dark matter abundance only if the particle spectrum is such that coannihilations or resonant annihilation channels are open.

To make the previous statement more quantitative, in this section we limit our discussion to neutralinos with a bino purity larger than 0.9. We find that:
\begin{enumerate}
\item Among the binos compatible with the {\em upper} WMAP bound, 91\% have at least one coannihilating partner\footnote{A particle is considered a coannihilating partner if the ratio $\Delta\Omega/\Omega$ between the difference of the relic density computed without taking into account coannihilations and that with coannihilation $\Omega$, over $\Omega$, is larger than 1\% (see next section for details).} and 15\% have a resonant annihilation cross section with the heavier Higgs boson mass $m_A\simeq 2\cdot m_\chi$. Only 2\% of binos do not feature any relic density suppression mechanism.
\item Binos which produce a relic density {\em within} the WMAP range have a coannihilating partner in 86\% of considered cases and a resonant cross section in 18\% of cases. 4\% of binos does not feature any relic density suppression mechanism.
\item The largest bino mass compatible with the WMAP bound, and such that neither coannihilations nor resonances are present, is around 160 GeV.
\end{enumerate}

It is therefore clear that in the large majority of cases, and for all masses larger than few hundreds GeV, binos can be dark matter candidates in the general MSSM {\em only if either coannihilations or resonances} are present. In this respect it is worthwhile to analyze in greater detail both relic density suppression mechanisms. 

\subsection{Coannihilations}\label{sec:coann}

\begin{figure*}[!t]
\begin{center}
\includegraphics[scale=0.6]{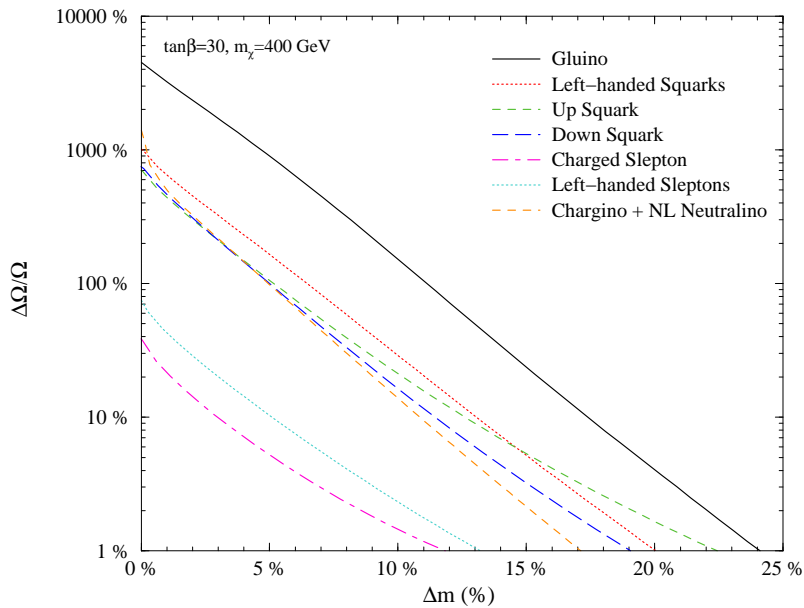}
\end{center}
\caption{\small \em Effectiveness of bino coannihilations in the MSSM: a plot of $\Delta\Omega/\Omega$ as a function of the mass splitting between a bino-line  neutralino and all possible coannihilating partners in the MSSM.}\label{fig:COMPARE}
\end{figure*}

The importance of {\em co-annihilations}, i.e. cases where the number density of a species depend not only on its own annihilation cross section but also on the effect of the annihilations of another species, was first recognized in the seminal papers of Ref.~\cite{Griest:1990kh,Edsjo:1997bg}. When the mass of the stable LSP is close to the mass of other particle species, the evolution of the number density of LSP's, $n_{\rm LSP}$, and that of the other $N$ particles, $n_{i=1,...,N}$, are tightly correlated, and must be tracked together while solving the Boltzmann equation
\begin{equation}
\frac{{\rm d}n}{{\rm d}t}=-3Hn-\langle\sigma_{\rm eff}v\rangle \left(n^2-n_{\rm eq}^2\right),
\end{equation} 
where the number density is now given by
\begin{equation}
n=n_{\rm LSP}+\sum_{i=1}^N n_i.
\end{equation}
In the equations above $H$ denotes the Hubble constant, $\langle\sigma_{\rm eff}v\rangle$ is the thermal average of the effective cross section times the relative velocity, and $n_{\rm eq}$ is the equilibrium number density, which, to a good approximation, may be taken to be equal to the Maxwell-Boltzmann thermal distribution. The main effects of taking into account coannihilations are two-folds: on the one hand one modifies the effective interaction cross section; on the other hand one alters the number of degrees of freedom which enters in the game. Therefore, if the cross section of the coannihilating partner is {\em much more} efficient than that of the stable species, the net result will be a {\em reduction} of the number density $n_{\rm LSP}$. In the opposite case, i.e. when the extra coannihilating degrees of freedom carry less efficient annihilations, the outcome can be an increase of the final asymptotic $n_{\rm LSP}$. Since, however, the coannihilating partners have typically non-zero electric or color charges, while neutralinos are neutral, this second effect is rather unusual (for exceptions see \cite{Edsjo:2003us} in the mSUGRA context and \cite{Profumo:2004wk} in the gluino coannihilation model).

Since the discovery of the generic mechanism of coannihilations, many dedicated studies have analyzed the impact of considering various coannihilating partners. In particular, within the framework of the {\em constrained} MSSM, the NLSP is found to be, in the low $m_0$ region, the lightest stau \cite{Ellis:1998kh,Ellis:1999mm,Nihei:2002sc,Edsjo:2003us}. In the focus point region of the CMSSM, instead, a non-trivial higgsino fraction may give rise to chargino and next-to-lightest neutralino coannihilation processes, see e.g.~\cite{Feng:2000gh,Birkedal-Hansen:2002sx,Bednyakov:2002js,Edsjo:2003us}. Again within the CMSSM, at large scalar trilinear couplings, it may well be that the next-to-LSP is the stop, whose coannihilations were considered  in \cite{Boehm:1999bj,Ellis:2001nx,Edsjo:2003us}. Minimal deviations from the assumed universality of scalar soft-breaking masses (SBM) have been shown to lead to other viable coannihilating partners: for instance, lowering the scalar SBM of the particles belonging to the ${\mathbf 5}$ representation of $SU(5)$ gives raise to coannihilations with the lightest bottom squark and with the tau sneutrino \cite{Profumo:2003em,Profumo:2003ki}. Relaxing the assumption of universality at the high energy (grand unification) scale in the Higgs sector may also give rise to sneutrino and other coannihilating partners which are not present in the CMSSM \cite{Ellis:2002wv,Ellis:2002iu}. Finally, in a recent analysis it has been shown that the strongest coannihilation processes in the MSSM are those with the {\em gluino}, and that the maximal bino mass is reached precisely in the gluino coannihilation tail \cite{Profumo:2004wk}.

The effect of including coannihilations in the computation of the relic density of neutralinos has been carried out for particular coannihilating partners, see e.g. in Ref.~\cite{Edsjo:2003us}. In \cite{Profumo:2004wk} we analyzed the relevant case of {\em bino} coannihilations with all possible partners, assuming, within an effective MSSM, that all relevant soft breaking masses are three times larger than the bino mass $m_1$, except for the particular coannihilating partner mass, which was taken to be close to $m_1$. The relevant parameter we used was the relative splitting between the bino mass and the mass of the coannihilating particle $m_{\widetilde P}$,
\begin{equation}
\Delta m\equiv\frac{m_{\widetilde P}-m_\chi}{m_\chi}
\end{equation}
The results are mostly independent of both the absolute size of $m_1$ and of the details of the spectrum of the other SUSY particles. Moreover, the dependence on $\tan\beta$ was found to be not critical.

\begin{table}[!t]
\begin{center}
\begin{tabular}{|c|c|c|c|}
\hline
{\bf Particle} & {\bf Symbol} & {$\mathbf \Delta m$} & {\bf \# of Particles}\\
\hline
Charged Slepton & $\widetilde l$ & 12\% & 6\\
Sneutrino & $\widetilde \nu$ & 13\% & 3\\
Up-type Squark & $\widetilde t$ & 21\%& 6\\
Down Squark & $\widetilde b$ & 20\% & 6\\
Gluino & $\widetilde g$ & 24\% & 1\\
NL-Neutralino(s) & $\chi^0$ & 17\% & 2\\
Chargino & $\chi^\pm$ & 17\% & 2\\
\hline
\end{tabular}
\end{center}
\caption{\small \em Mass splitting within which coannihilations with different particles are effective.}
\label{tab:cuts}
\end{table}

We show in Fig.~\ref{fig:COMPARE} the relative difference in relic density without and with coannihilations \cite{Edsjo:2003us}
\begin{equation}
\frac{\Delta\Omega}{\Omega}\equiv\frac{\Omega^{\rm no\ coann}_\chi-\Omega^{\rm coann}_\chi}{\Omega^{\rm coann}_\chi}.
\end{equation}
The lower limit on $\Delta\Omega/\Omega$ has been set to 1\%, since this is the typical numerical accuracy of the numerical packages we employed to carry out the computations \cite{Edsjo:2003us,Belanger:2001fz}. The plot was taken at a neutralino mass $m_\chi=400$ GeV and at $\tan\beta=30$.

From the plot we clearly deduce that gluino coannihilations are the strongest possible bino coannihilation processes in the MSSM, as it might be expected considering the strong gluino-gluino annihilation cross section and the number of final SM states, much larger than that of squarks. Binos can also annihilate with a quasi degenerate wino without altering its bino purity (the same does not hold true for the higgsino, i.e. in case $m_1\simeq \mu$, due to the neutralino mass matrix structure), and therefore undergo chargino and next-to-lightest neutralino coannihilations. These processes are found to be of the same order of magnitude of those involving up and down squarks (or two of them at the same time). Much more suppressed are, instead, slepton coannihilations.

Fig.~\ref{fig:COMPARE} allows us to make a quantitative statement about models where coannihilations are or not relevant in reducing the relic abundance of binos to acceptable levels. We will declare that a model has a coannihilating partner $\widetilde P$ if the mass splitting of that partner from the bino mass, $\Delta m_{\widetilde P}$, is such that the corresponding $\Delta\Omega/\Omega$ is larger than 1\%. In this respect, we show in Tab.~\ref{tab:cuts} the cuts on the mass splitting corresponding to all possible bino coannihilating partners, as well as the symbols we will use in the following figures.

\begin{figure*}[!b]
\begin{center}
\begin{tabular}{ccc}
\includegraphics[scale=0.75]{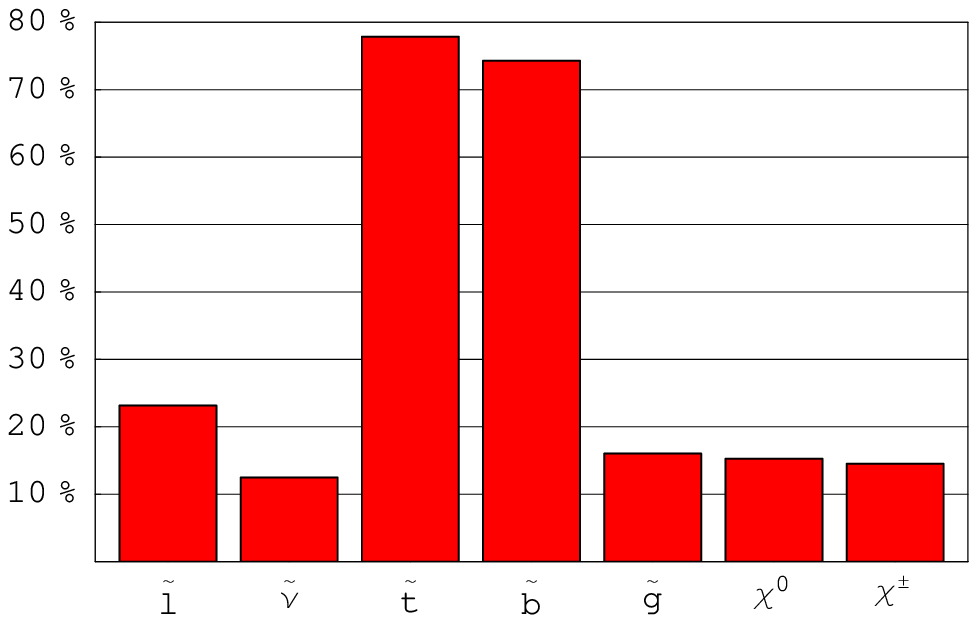} & \hspace{0.5cm} &\includegraphics[scale=0.75]{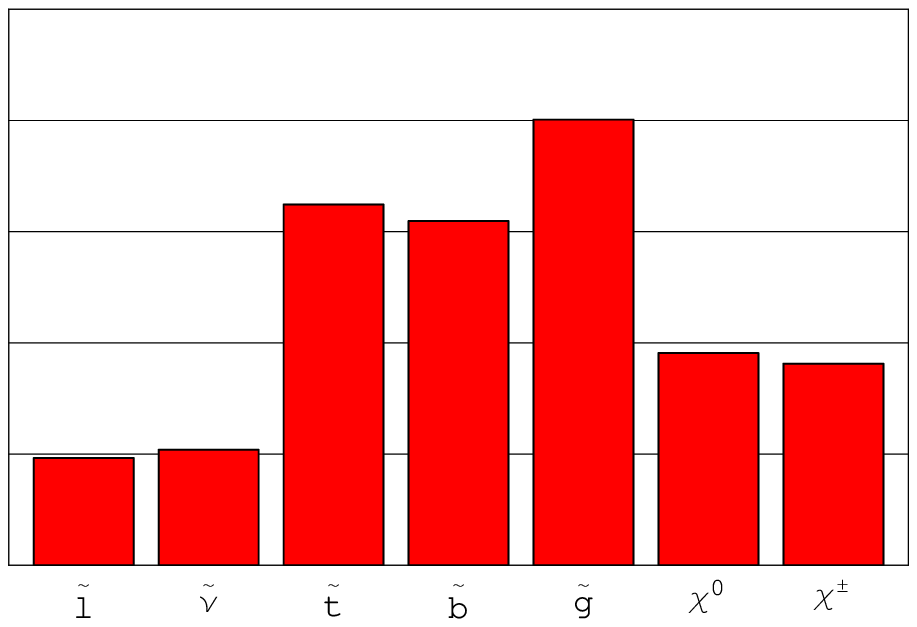}\\
({\em a\ }) && ({\em b\ })\\
\end{tabular}
\end{center}
\caption{\small \em Bino-coannihilations in the MSSM. In $(a)$, we show, for each possible coannihilating partner, the percentage of viable models with bino-like neutralinos in which coannihilations with one of those partners take place. In $(b)$ we have normalized the data from $(a)$ dividing by the number of particles in each class (e.g. by six for $\widetilde l$, by three for $\widetilde \nu$, etc. ).}\label{fig:coan}
\end{figure*}

We then consider all the models in our scan featuring a bino-like LSP with a purity larger than 90\%, and whose relic abundance falls {\em within} the WMAP bounds, and look for the number of models exhibiting a given coannihilation channel. Our results are summarized, with the same symbols as in Tab.~\ref{tab:cuts}, in Fig.~\ref{fig:coan} ({\em a\ }). We see that in most models with the correct neutralino relic density a squark coannihilation is present: this is mainly due to the fact that squark coannihilations are strong, but also to the number of different particles falling into this category (e.g. 6 up-type squarks). Therefore, to single out the relevance of coannihilation processes with a given species, we divide the number of models characterized by the occurrence of a given coannihilating partner by the number of possible coannihilating partners of that given species. We show our results in Fig.~\ref{fig:coan} ({\em b\ }): remarkably, we obtain the same hierarchy of coannihilation efficiency outlined in Fig.~\ref{fig:COMPARE}: gluino coannihilations are the strongest one, followed by squark coannihilations, chargino and NL-neutralino and finally sleptons.

\begin{figure*}[!t]
\begin{center}
\begin{tabular}{cc}
\includegraphics[scale=0.65]{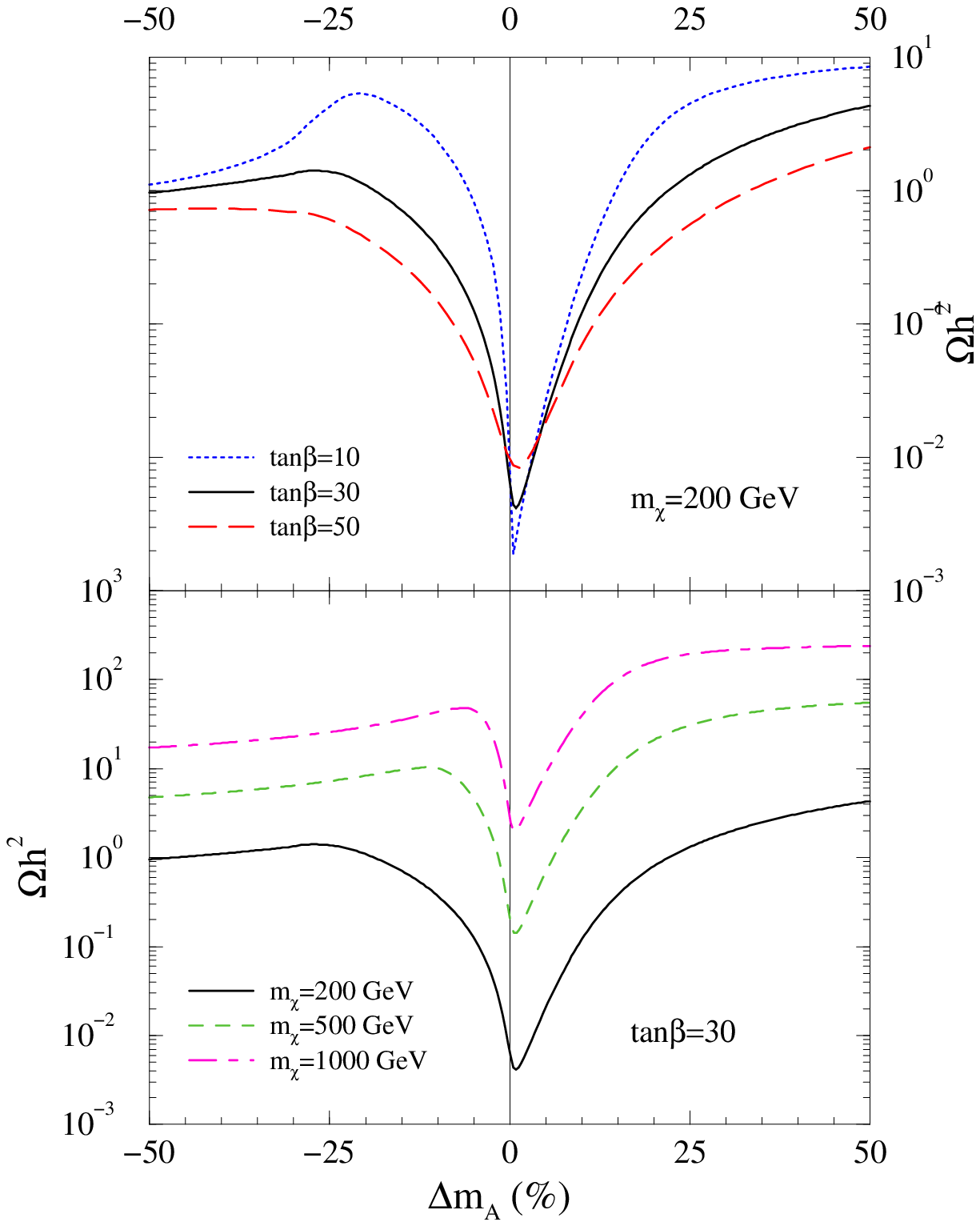} 
& 
\includegraphics[scale=0.75]{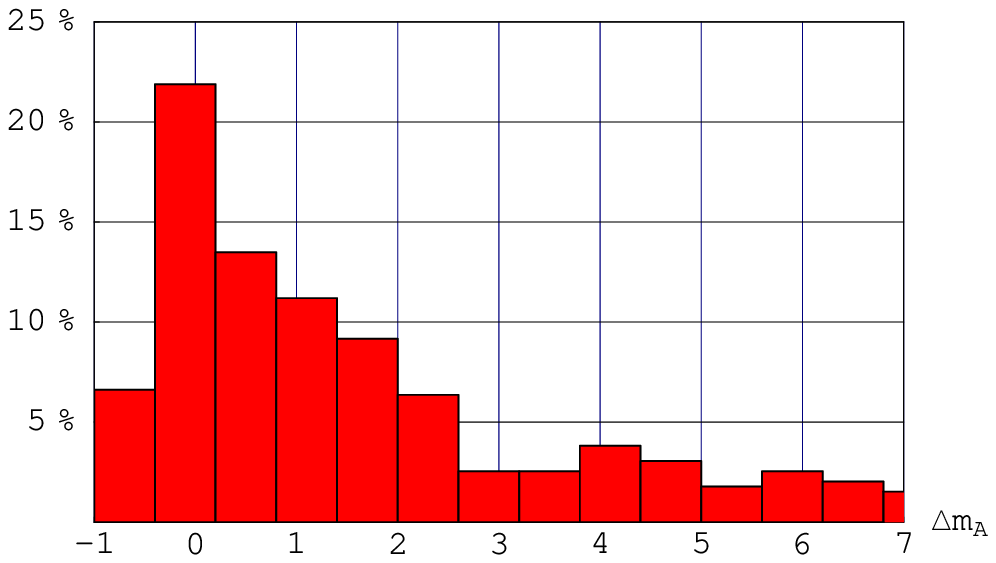}\\
({\em a\ }) & ({\em b\ })\\
\end{tabular}
\end{center}
\caption{\small \em The relic density in the presence of a resonant annihilation cross section, as a function of $\Delta m_A$, for different values of $\tan\beta$ and different neutralino masses $(a)$. In $(b)$ we display an histogram of $\Delta m_A$ for all viable models with a bino-like neutralino and $\Omega h^2$ in the WMAP range.}\label{fig:FUNN}
\end{figure*}

\subsection{Resonances}

The computation of the relic density of a thermally produced particle $\chi$ needs particular care when the mass of $\chi$ lies close to half the mass of another particle $P$ in which $\chi$ can annihilate through an $s$-channel $P$ exchange. In this case, in fact, the propagator of the $s$-channel diagram has a resonant behavior. This is the case, for instance, in the MSSM for the heavy $CP$-odd $A$ and $CP$-even $H$ Higgs bosons\footnote{Another possibility is a resonance with the lightest $CP$-even Higgs, which would however require a neutralino mass below 60-70 GeV: this case has therefore not been considered in the present study.}. The calculation of the thermally averaged cross section for the resonant $\chi\chi$ annihilation cross section needs the thermally averaged treatment of a cross section of the Breit-Wigner form \cite{resonance}, see e.g. \cite{Gondolo:1990dk}. 

The effects of a heavy Higgs resonance on the annihilation cross section of neutralinos depends on three inputs: first, the mass splitting between the annihilating particle $\chi$ and half the mass of the resonantly exchanged particle $A$, which we quantify through the parameter
\begin{equation}
\Delta m_A\equiv \frac{m_A-2\cdot m_\chi}{2\cdot m_\chi};
\end{equation}
second,  the total decay width of the resonance $\Gamma_A$ and  third its mass $m_A$ \cite{Gondolo:1990dk}. Finally, the ratio of the naively computed cross section and the correct one depends on the relevant $\chi\chi A$ couplings and on the couplings of the resonance to the final particle states. For recent studies dedicated to the effects of resonant annihilation cross sections see \cite{Lahanas:1999uy}.  

We show in Fig.~\ref{fig:FUNN} the overall non-trivial combination of the effects of these input parameters. In the upper panel we show the dependence of the neutralino relic density $\Omega_\chi h^2$ on $\Delta m_A$ at fixed $m_\chi=200$ GeV for three different values of $\tan\beta$, while in the lower panel we fix $\tan\beta=30$ and analyze three different $m_\chi$=200, 500 and 1000 GeV. The plot clearly highlights the effect of the resonance around $\Delta m_A=0$, and in particular the results of the {\em thermal averaging}: in fact, the maximal effects are at {\em positive} $\Delta m_A\approx 1$\%, since particles with mass $m_\chi\lesssim m_A/2$ in general have larger thermal energy contributions than those  with  $m_\chi\gtrsim m_A/2$. 

In the upper panel we see the effect of a broader resonance (the width of $m_A$ grows with $\tan\beta$) on the relic density: for larger $\tan\beta=50$ resonance effects are effective for a larger $\Delta m_A$ range. In the lower panel, instead, it is shown how larger masses reduce the efficiency of resonance effects and shrink the $\Delta m_A$ range where resonance effects are relevant.

Resonances have drastic consequences on the cosmological constraints on the MSSM parameter space, mainly when the $\chi\chi$ annihilation cross section is relatively low, as it is the case for a bino-like LSP. The plot on the right shows the overall statistical effects of the $A$ pole resonance showing the fraction of binos in a given bin of $\Delta m_A$. We clearly see that the privileged values are those around $\Delta m_A=0$, as expected. 


\section{Direct and Indirect DM detection}\label{sec:det}

The present section is devoted to the discussion of direct and indirect dark matter detection signals in the general MSSM (for recent reviews on the subject of dark matter detection see \cite{Munoz:2003gx}). The issue of the {\em comparison} between different search strategies, and of the complementarity among them, has since long been the subject of various investigations (see e.g. \cite{earlycomparison}). In the present context we propose an approach which features three major novelties:

\begin{enumerate}
\item We adopt self-consistent halo-models, along the lines of Ref.~\cite{pierohalos}, where two extreme benchmark scenarios for the dark matter distribution in the Milky Way have been outlined, both being soundly motivated from the point of view of structure formation, and consistent with all observational data available and with numerical simulations. The two models represent two different back-reaction mechanisms of the baryon infall on the dark matter distribution, giving rise to a non-cuspy profile (the {\em Burkert} profile) and to a steeply cuspy profile ({\em Adiabatically contracted} profile). The Burkert profile is a  {\em conservative} halo-model  of the ``cored'' type: the central cusp gets smoothed out through large angular momentum transfer between the baryonic and the dark matter components. In the adiabatically contracted profile, {\em optimistic} from the point of view of dark matter detection, the starting dark matter distribution (assumed to be the CDM profile of Ref.~\cite{NFW}) undergoes an adiabatic contraction with no net angular momentum transfer: The CDM cusp therefore gets increased, approaching a rather steep density profile towards the Galactic center \cite{pierohalos}. We stress that both profiles satisfy all dynamical constraints, and that the velocity distributions are computed self-consistently. Moreover, they represent two extreme cases: it is reasonable to assume that the actual halo profile of the Milky Way lies in between these two cases. In this respect, our results can be regarded as representative lower and upper limits on SUSY dark matter detection rates. Let us stress that, in order to compare different search strategies, it is mandatory to have consistent profiles for dark matter densities and velocities, and both must be compatible with observational data.
\item We make use of a statistical analysis based on {\em Visibility Ratios}, i.e. signal-to-sensitivity ratios, which allows a transparent comparison between different methods and between current and projected sensitivities. Our aim is to estimate the statistical relevance of a given search strategy. 
We claim that this approach may be more quantitative and helpful, in the evaluation of the perspectives of detection in the most general SUSY setup, than usual scatter plots, which hardly bring any quantitative statistical information. We will however also make use of scatter plots in order to show correlations among different quantities, and to outline the dependence of detection rates on the neutralino mass, composition and relic abundance.
\item In view of recent results about the occurrence of relic density enhancement mechanisms in modified cosmological contexts \cite{quint,Profumo:2003hq,shear, Profumo:2004ex,Catena:2004ba}, or through non-thermal neutralino production \cite{non-therm}, we will not apply any rescaling procedure in the computation of detection rates. Models with a low relic abundance may well be responsible for the whole inferred amount of dark matter in the Universe, and provide, thanks to large neutralino annihilation amplitudes, significant indirect detection rates. We will nevertheless always single out, for definiteness, the case of models which provide the required neutralino abundance in the standard thermal cosmological scenario.  
\end{enumerate}

\subsection{Spin Dependent and Spin Independent Direct Detection}\label{sec:dirdet}

\begin{figure*}[!t]
\begin{center}
\begin{tabular}{cc}
\includegraphics[scale=0.48]{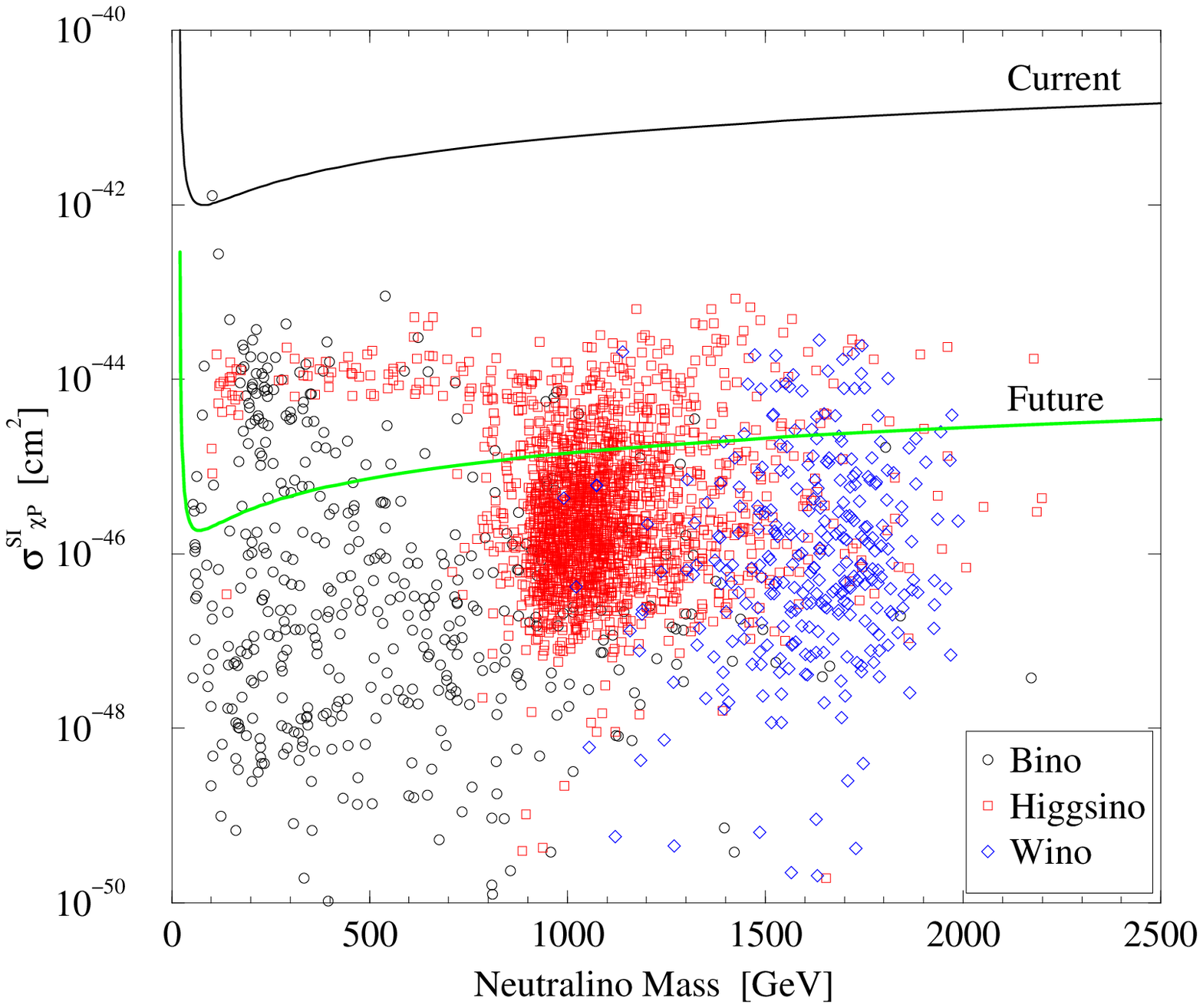} & \includegraphics[scale=0.48]{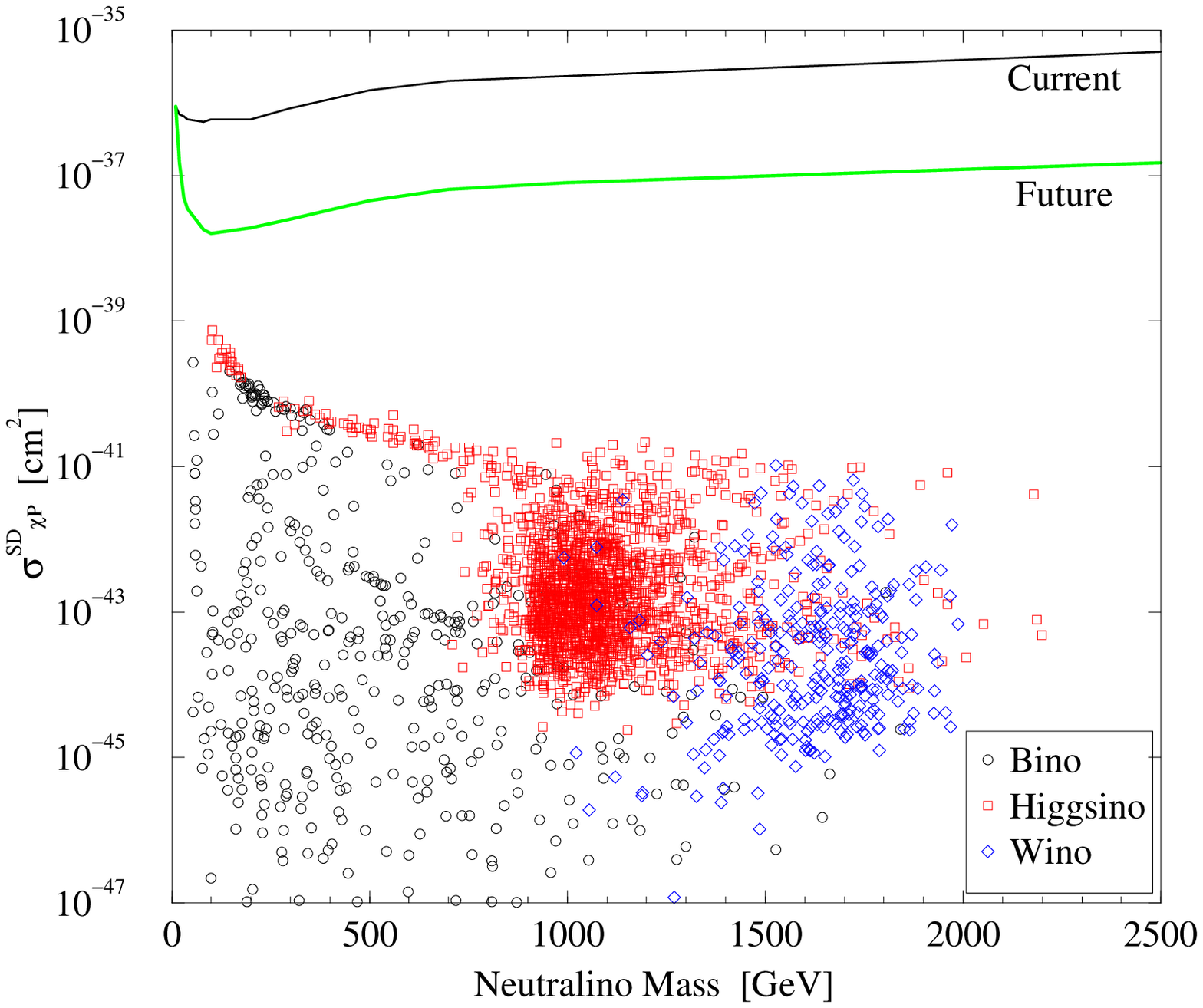}\\
({\em a\ }) & ({\em b\ })\\
\end{tabular}
\end{center}
\caption{\small \em A scatter plot of the Spin Independent (a) and Spin Dependent (b) neutralino-proton scattering cross sections for models whose relic density lies {\em within the WMAP range}. Black circles represent models in which the lightest neutralino is predominantly a Bino, red squares represent Higgsinos and Blue diamonds Winos. The black lines indicate the current exclusion limits \cite{edel}, while green lines the projected exclusion limits at future experiments \cite{xenon}.}\label{fig:DIRDET}
\end{figure*}

We show in Fig.~\ref{fig:DIRDET} the results of the scan  for direct detection experiments in models with a relic density within the WMAP range. On the left we plot the {\em spin-independent neutralino-proton cross section} (results for the neutralino-neutron cross sections are very similar) including, in black, the current limits from the Edelweiss experiment \cite{edel} and, in green, the projected future sensitivity of the Xenon 1-ton facility \cite{xenon}, computed for the  Adiabatically contracted halo profile\footnote{In the case of direct detection and neutrino telescopes, the dependence on the halo profile, which is merely local, is rather mild, see \cite{pierohalos}.}. The  code for the scatter plot is as follows: black circles refer to  bino-like neutralinos, red squares to higgsino-like and blue diamonds to wino-like.

\begin{figure*}[!t]
\begin{center}
\includegraphics[scale=0.8]{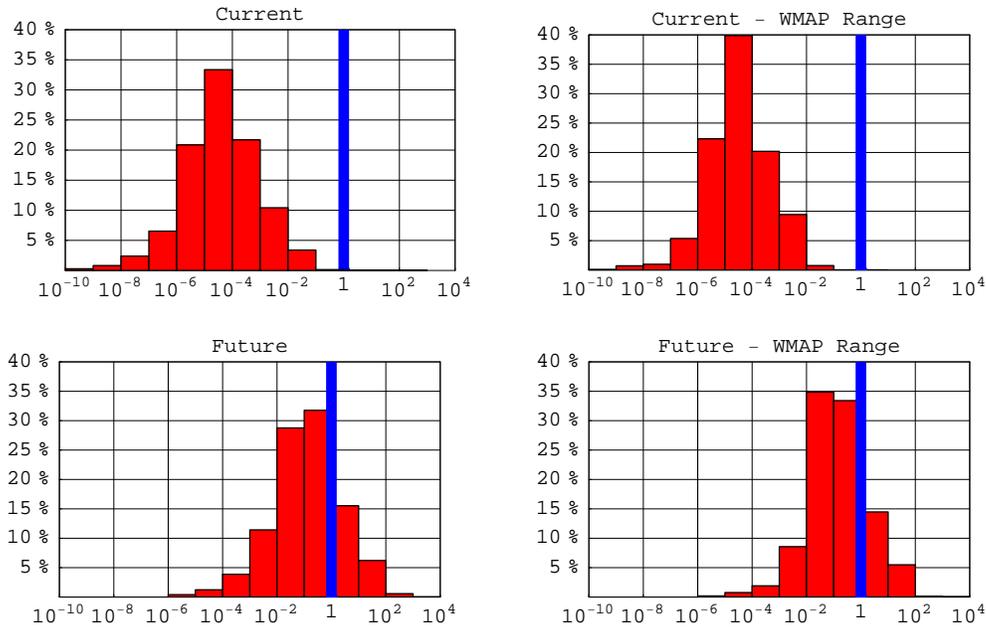}\\
\end{center}
\caption{\small \em A statistical analysis of the Visibility Ratio (signal-to-sensitivity) for the {\em Spin Independent neutralino-proton cross section} for all models with a sufficiently low relic abundance  (plots to the left) and for models with a relic abundance in the WMAP range (plots to the right). The two upper plots refer to the current exclusion limits, as in Fig.~\ref{fig:DIRDET} (a), black line, while the two lower plots to the future exclusion limits, Fig.~\ref{fig:DIRDET} (a), green line.}\label{fig:DIRDEThisto}
\end{figure*}

The right part of the figure shows instead our results for the {\em spin-dependent} neutralino-proton cross section, with current constraints from various existing experiments indicated with a black line \cite{ref:dmsens}, and the future projected NAIAD sensitivity in green \cite{NaI}. As evident from the figure,  spin-dependent searches always feature an experimental sensitivity which is unable to probe any viable SUSY model in our scan.

Fig.~\ref{fig:DIRDEThisto} collects the statistical analysis of our results for {\em spin-independent searches}, as a function of visibility ratios,  for the adiabatic halo profile. The histograms to the left show the cumulative results for all models, while those to the right refer to models within the WMAP range (hence those shown in the scatter plots of Fig.~\ref{fig:DIRDET}). The two upper histograms refer to the current limits (those indicated with black lines in Fig.~\ref{fig:DIRDET}), and the two lower ones to future exclusion limits (green lines in Fig.~\ref{fig:DIRDET}). The vertical blue lines indicate the visibility threshold: models lying to the right of the blue line are  {\em above} current or future sensitivities, those to the left currently give no (or will not give, at future experimental facilities) detectable signals. While a negligible fraction of models lie above current exclusion limits, approximately 20\% of models (in the WMAP range or with lower relic abundances) will be accessible to Xenon 1-ton \cite{xenon}, or to similar detection experiments.

A few comments are in order: first, spin dependent rates are found to be at least one order of magnitude below future sensitivities\footnote{The same applies also for low relic density models, which we do not include in Fig.~\ref{fig:DIRDET}.}. {\em The potential of spin dependent searches is then  statistically negligible.} Second,  current exclusion limits from spin-independent searches are still far from probing a significant portion of the general MSSM parameter space; on the other hand, future experiments will be able to probe $20\%$ of viable models. The scatter plots also show the mass clustering of pure higgsinos and of winos around, respectively, 1 TeV and 1.6 TeV: the dependence of the neutralino-proton scattering cross section on the details of the SUSY spectrum, particularly concerning the Higgs sector (the sign of the $\mu$ parameter and the masses of the $CP$-even $h$ and $H$ neutral Higgses), and, to a less extent, the squark sector yield a scatter in $\sigma^{\rm SI}_{\chi \rm P}$ which can well be as large as {\em four orders of magnitude}.

\subsection{Neutralino Induced Muon Fluxes}

\begin{figure*}[!t]
\begin{center}
\begin{tabular}{cc}
\includegraphics[scale=0.48]{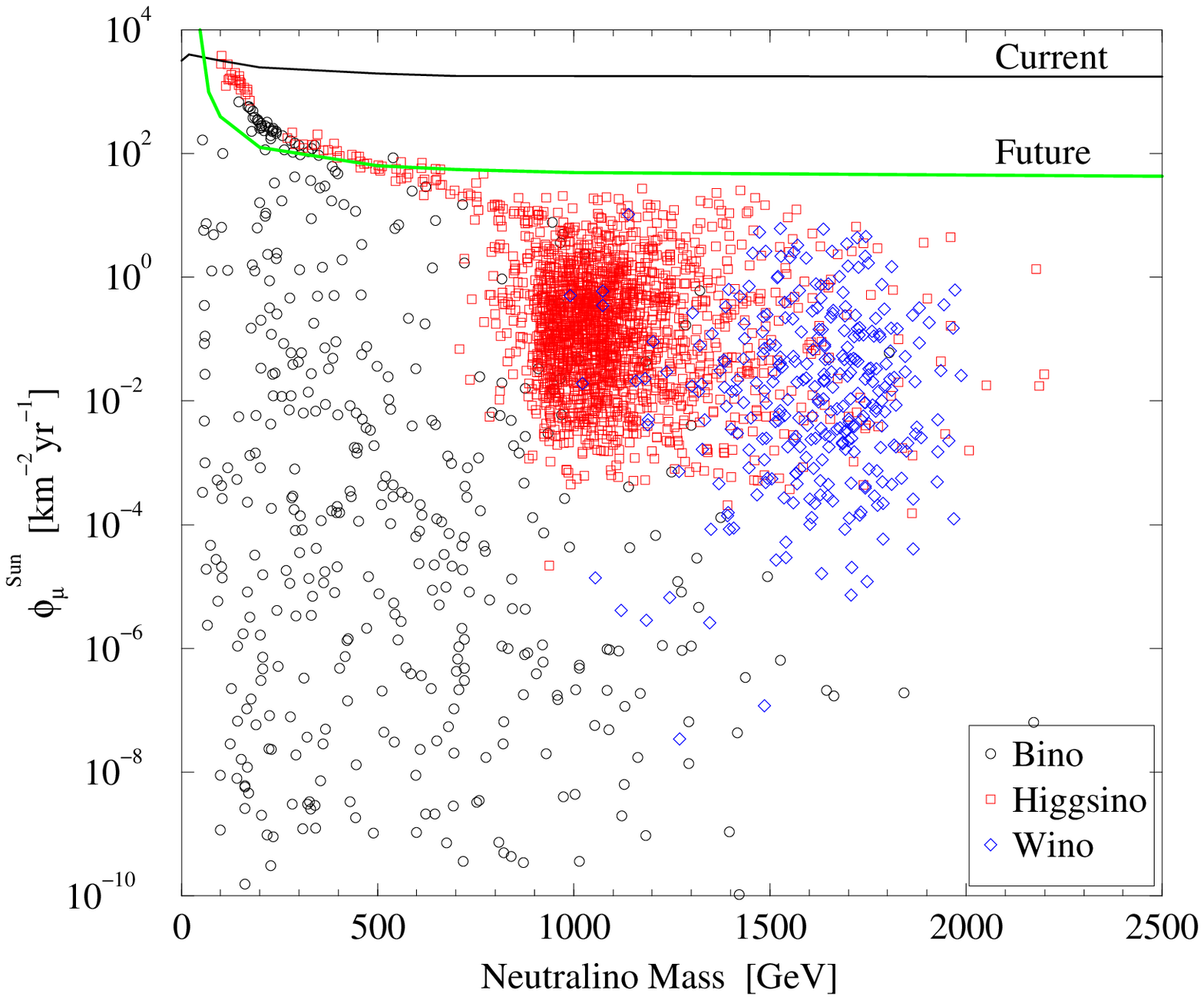} & \includegraphics[scale=0.48]{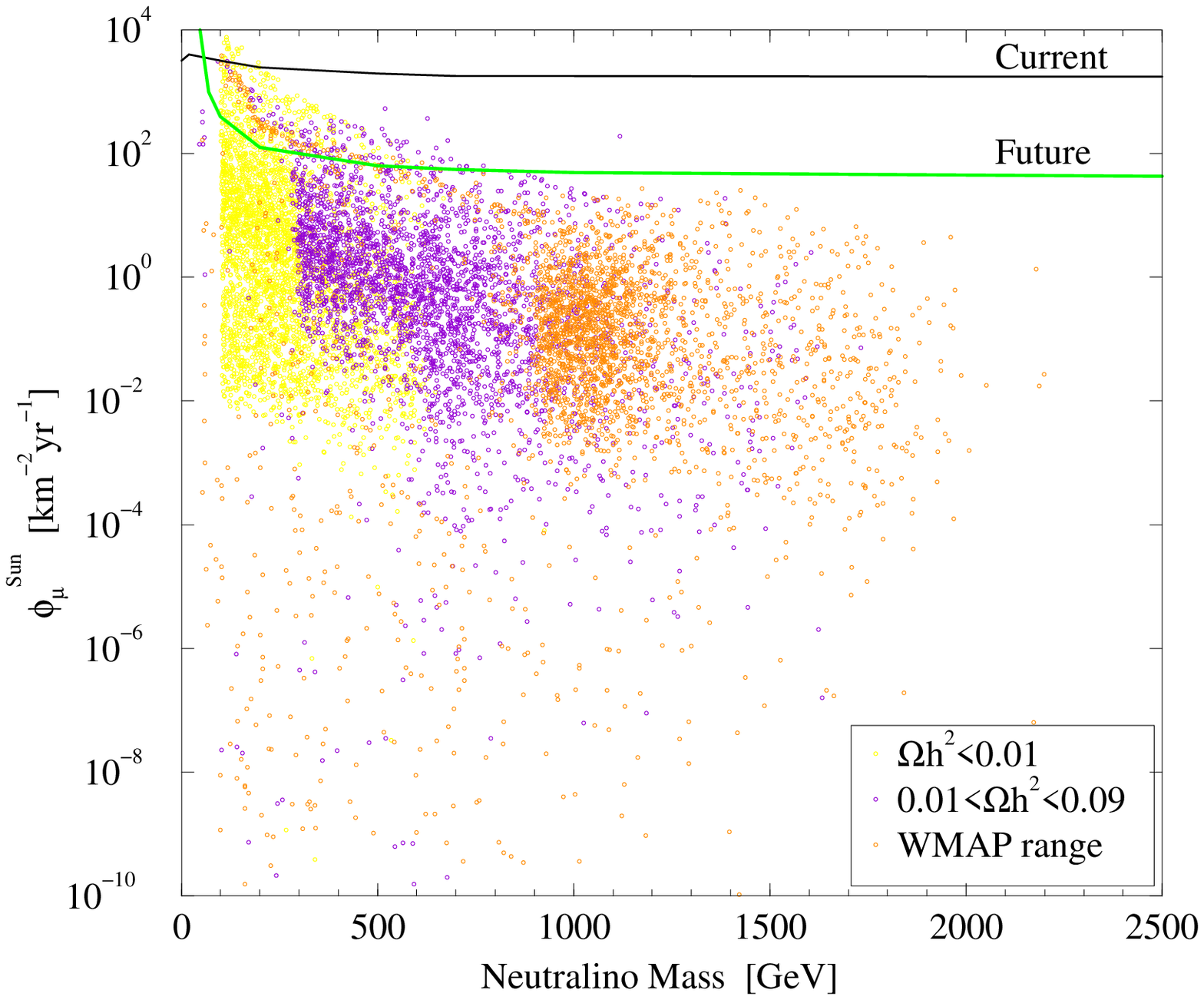}\\
({\em a\ }) & ({\em b\ })\\
\end{tabular}
\end{center}
\caption{\small \em Scatter plots of the expected muon flux at neutrino telescopes from neutralino annihilations in the center of the Sun, above a 1 GeV threshold. Models in (a) have relic abundances in the WMAP range, and are grouped according to the bino, higgsino and wino content of the lightest neutralino. In (b) we show instead a selection of all models grouped by relic abundance. We also show, respectively with a black and a green line, current and future experimental sensitivity limits.}\label{fig:NEUTTLSC}
\end{figure*}
\begin{figure*}[!t]
\begin{center}
\includegraphics[scale=0.8]{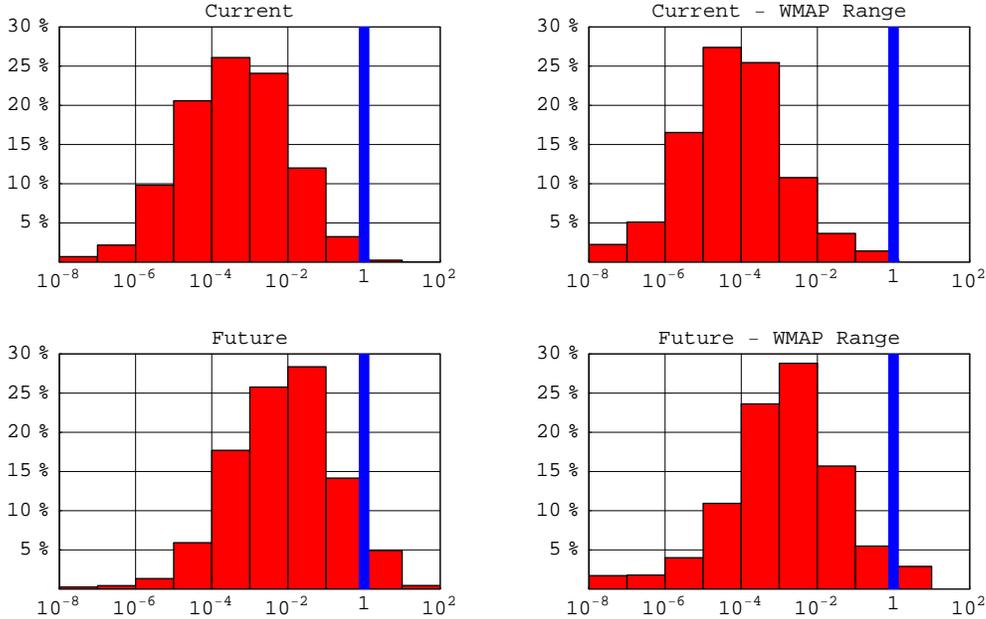}\\
\end{center}
\caption{\small \em A statistical analysis of the Visibility Ratio (signal-to-sensitivity) for the {\em Muon flux from the Sun at Neutrino Telescopes} for all models with a sufficiently low relic abundance  (plots to the left) and for models with a relic abundance in the WMAP range (plots to the right). The two upper plots refer to the current exclusion limits, as in Fig.~\ref{fig:NEUTTLSC} (a), black lines, while the two lower plots to the future exclusion limits, Fig.~\ref{fig:NEUTTLSC}, green lines.}\label{fig:NThisto}
\end{figure*}

It is often stated that one of the best indirect neutralino search strategies is to look for neutrinos produced by neutralino annihilations in the center of the Sun or of the Earth. Neutralinos from the galactic halo may get trapped into these astrophysical bodies, and begin to sink into the center, where the density enhancement produces annihilations into SM particles, among which neutrinos, the only particles which may later be detected. Neutrino telescopes with ${\rm Km}^2$ size may be able to single out the neutrino flux from the center of those celestial bodies, and distinguish it from the unavoidable neutrino background from cosmic rays interactions with the atmosphere. A crucial quantity in this game is the balance between the capture rate and the annihilation rate: if this is an equilibrium process, then the signal is at his maximum; otherwise large suppressions may occur. The equilibrium time scale inside the Sun is in most cases much smaller compared to that in the Earth, and though, depending on the SUSY model, equilibrium may not be reached even in the Sun, it very rarely occurs in the case of the Earth. As a result, the flux of muon neutrinos from the Earth is, in most cases, far below current and future sensitivities; for this reason, we do not show here results for this detection channel. 

We show in Fig.~\ref{fig:NEUTTLSC} that the flux from the Sun may be large enough to be detectable at future experiments (we will use here the future sensitivity prospects for the IceCube experiment \cite{icecube}). A few models are even already excluded by current SuperKamiokande data \cite{superKlimit}. It goes without saying that models with a larger annihilation cross section give larger rates: this point is clarified in the right panel of Fig.~\ref{fig:NEUTTLSC}, where we include also low relic density models. Interestingly, we find that, in any case, {\em DM searches at neutrino telescopes will not probe neutralino masses larger than about 750 }GeV. Notice that the overall gross features for spin-dependent rates and for the muon flux from the Sun are rather similar to each other: this does not come as a surprise, since the capture rate into the Sun, mainly composed of nuclei with spin different from zero, depends in fact on the spin-dependent neutralino-nucleon cross section.

The suitable neutralino candidate whose relic abundance lies within the WMAP range, and which will be detectable at neutrino telescopes, is a composite bino-higgsino state with a large enough spin-dependent cross section, and a mass below half a TeV. Models at larger masses (as pure Higgsinos and Winos with relic abundances within the WMAP range) will not be probed at future experiments.

The statistical summary of our scan is shown in the histograms of Fig.~\ref{fig:NThisto}. Once again, current experiments only probe a marginal fraction of models, particularly if lying into the WMAP preferred range (histograms to the right). On the other hand, future experiments will be able to probe from 3 to 5\% of the viable models.

\subsection{Correlating Direct Searches and Muon Fluxes}

\begin{figure*}[!b]
\begin{center}
\begin{tabular}{cc}
\includegraphics[scale=0.5]{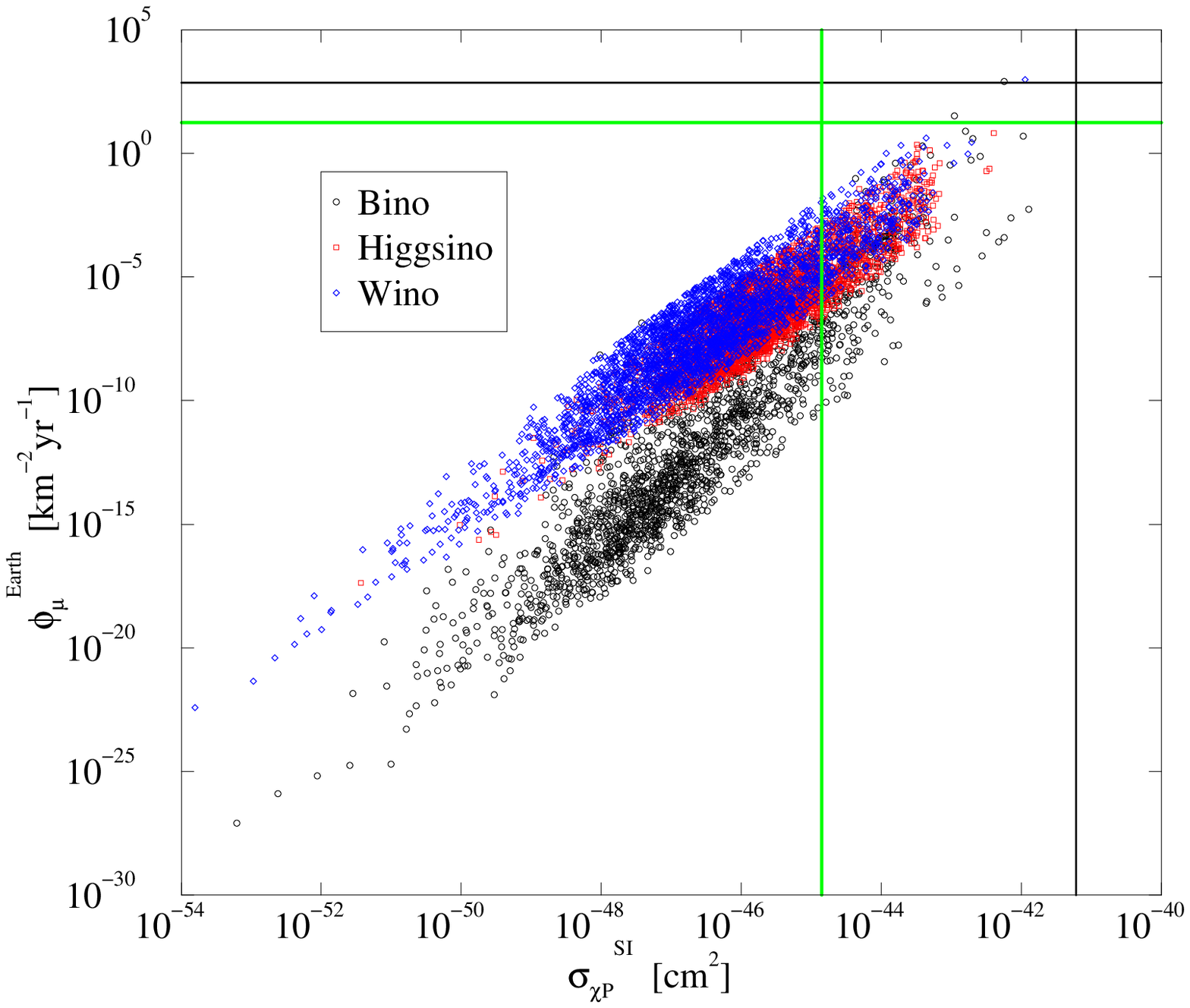} & \includegraphics[scale=0.5]{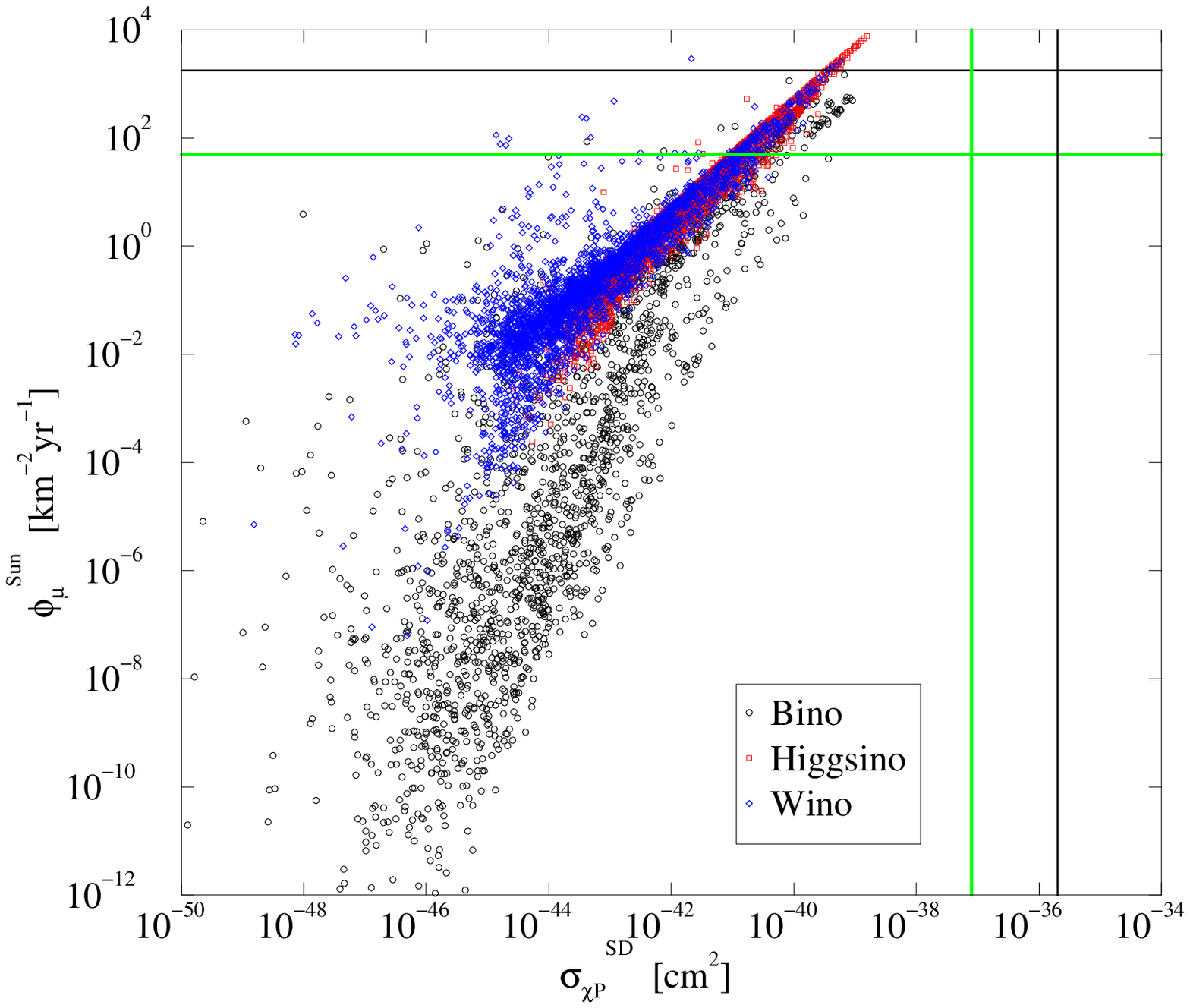}\\
({\em a\ }) & ({\em b\ })\\
\end{tabular}
\end{center}
\caption{\small \em The correlation between the muon flux from the Earth and the Spin Independent neutralino-proton scattering cross section (a) and the correlation between the muon flux from the Sun and the Spin Dependent neutralino-proton scattering cross section (b). We show here a reduced sample of all models considered (including models with relic density below the WMAP range). Black circles represent models in which the lightest neutralino is predominantly a Bino, red squares represent Higgsinos and Blue diamonds Winos. As guidelines, we also include green and black lines representing future and current sensitivities at a neutralino mass of 1000 GeV.}\label{fig:CORRDIR}
\end{figure*}

In this section we analyze the correlation between neutralino direct searches and indirect searches at neutrino telescopes. In particular, as outlined above, we correlate, in Fig.~\ref{fig:CORRDIR}, spin-dependent neutralino-proton rates with the muon flux from the Sun, and the scalar neutralino-proton cross section with the rate of muons from the Earth (mainly composed by spin-less nuclei). As guidelines, we also include current and future sensitivities, at a putative neutralino mass of 1 TeV: models above (or at the right) the horizontal (respectively vertical) lines will be, or currently are, above projected, or current, sensitivity. The same color code as in Fig.~\ref{fig:DIRDET} has been used, although we plot here also models with low relic abundances.

As a first remark, we point out the well known complementarity between direct and indirect searches: models which are not visible at spin-dependent searches will be accessible at indirect searches in the muon-from-the-Sun channel, and, vice-versa, models with a too low muon flux from the Earth may well be above visibility threshold at direct spin-independent searches.

The correlation between indirect searches at neutrino telescopes and direct detection experiments is clearly visible, though some comments are in order. First, binos always tend to have a smaller muon flux, at the same neutralino-nucleon cross section, than higgsinos or winos. This fact is due to the pair annihilation rate, which in the case of binos is typically more suppressed than for winos or higgsinos; the sufficiently low relic abundance of binos has been shown to occur in most cases thanks to {\em coannihilation processes}, which are however not present for pair annihilations in the potential wells of the Sun or of the Earth. Secondly, the correlation between $\phi_\mu^{\rm Earth}$ and $\sigma_{\chi \rm P}^{SI}$, though clearly present, is scattered over at least four orders of magnitude in $\phi_\mu^{\rm Earth}$ at a given $\sigma_{\chi \rm P}^{SI}$, for each neutralino type. Various factors contribute to this spread: the mentioned effect due to the pair annihilation rate, the details of the annihilation-capture interplay and possible enhancements due to kinematical effects for particular values of the neutralino mass. 

We recall that, if equilibrium is reached, both $\phi_\mu^{\rm Earth}$ and $\sigma_{\chi \rm P}^{SI}$ scale as the squared of the nucleon matrix element of the effective Lagrangian for the scalar neutralino-nucleus interaction $|\langle\mathcal L_{\rm sc}\rangle|^2$, while if it is not then \cite{earlycomparison}
\begin{equation}
\phi_\mu^{\rm Earth}\propto |\langle\mathcal L_{\rm sc}\rangle|^4 \langle\sigma_{\rm ann} v\rangle_0
\end{equation}
$\langle\sigma_{\rm ann} v\rangle_0$ being the neutralino annihilation times the relative velocity in the zero velocity limit. In the case of the  $\phi_\mu^{\rm Sun}$ - $\sigma_{\chi \rm P}^{SD}$ correlation, we notice that large fluxes, corresponding to cases where annihilation and capture are in equilibrium, tend to have, as expected, an extremely strong correlation, which is lost when the signal is weaker, once again because equilibrium is not reached, and the dependence on $\langle\sigma_{\rm ann} v\rangle_0$ again enters into the game. In particular, this is the case for binos, where coannihilation effects with a large variety of partners can drastically affect the actual neutralino annihilation rate with respect to what expected from cosmological abundance arguments: a coannihilating bino can in fact produce a sufficiently reduced relic abundance though featuring a large $\langle\sigma_{\rm ann} v\rangle_0$.

\subsection{Antiprotons, Positrons and Gamma Ray Searches}

Antimatter searches (see e.g. Ref.~\cite{Bergstrom:1999jc}) have been recently shown to be an appealing search strategy for models with large annihilation rates, and therefore low relic abundances in the standard cosmological scenario \cite{Profumo:2004ty}. Large sources of uncertainties arise however in this context, mainly due to the computation of the secondary antiproton flux (the {\em background}), to the antimatter propagation in the Galaxy and in the Solar System \cite{Donato:2003xg}, and to uncertainties in the dark matter halo distribution in the Milky Way \cite{pierohalos}. 

We will here mainly refer to the analysis carried out in Ref.~\cite{Profumo:2004ty}. The primary source diffusion in the Galaxy is modeled with an effective two-dimensional diffusion setup in the steady state approximation; the solar modulation effects are taken into account with the analytical one parameter force-field approximation of Ref.~\cite{gleesonaxford}. The secondary antimatter fluxes are computed with the {\tt Galprop} package, and give an excellent fit to the currently available data, respectively with a reduced $\chi^2=0.82$ for antiprotons and 0.95 for positrons.

In Ref.~\cite{Profumo:2004ty} a given SUSY model was ruled out if the sum of the primary and secondary antimatter fluxes gave a statistically unacceptable $\chi^2$. For future experiments, a new parameter $I_\phi$ was proposed, basically defined as the integral over antimatter kinetic energies of the ratio of the squared of the signal over the background. This quantity could then be compared against quantities related to the particular experimental facility at given data acquisition time scales, providing an assessment of the visibility of the given SUSY model. Here, for computational ease, we will take one representative energy bin, both at current and future antimatter search experiments, and compare the primary antimatter flux with the (current or projected) experimental sensitivity. Since the background is consistent with available data, we will indicate as visible those models where the antimatter flux of supersymmetric origin $\phi^{\bar{a}_{\rm SUSY}}$ gives a contribution which could have been, or which will be detected, i.e.
\begin{equation}
\phi^{\bar{a}_{\rm SUSY}}\gtrsim 2\ \sigma_{\rm exp},
\end{equation}
where $\sigma_{\rm exp}$ indicates the current, or projected, sensitivity in the given energy bin. We compared, for a sample of models, the method we use here with the fully statistical accurate strategy of Ref.~\cite{Profumo:2004ty}, and we verified that the exclusion limits found were consistent with each other.

\begin{figure*}[!b]
\begin{center}
\begin{tabular}{cc}
\includegraphics[scale=0.5]{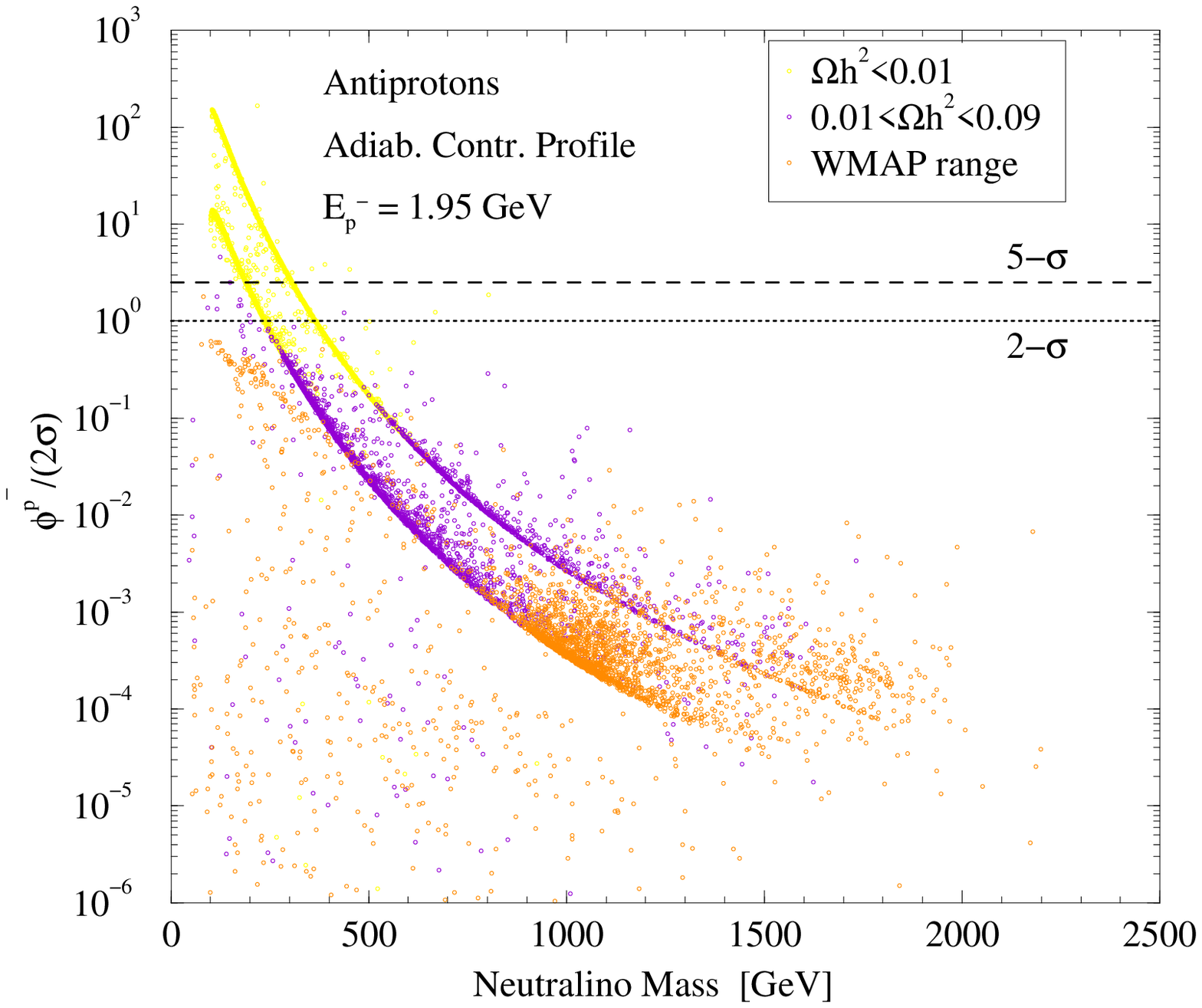} & \includegraphics[scale=0.5]{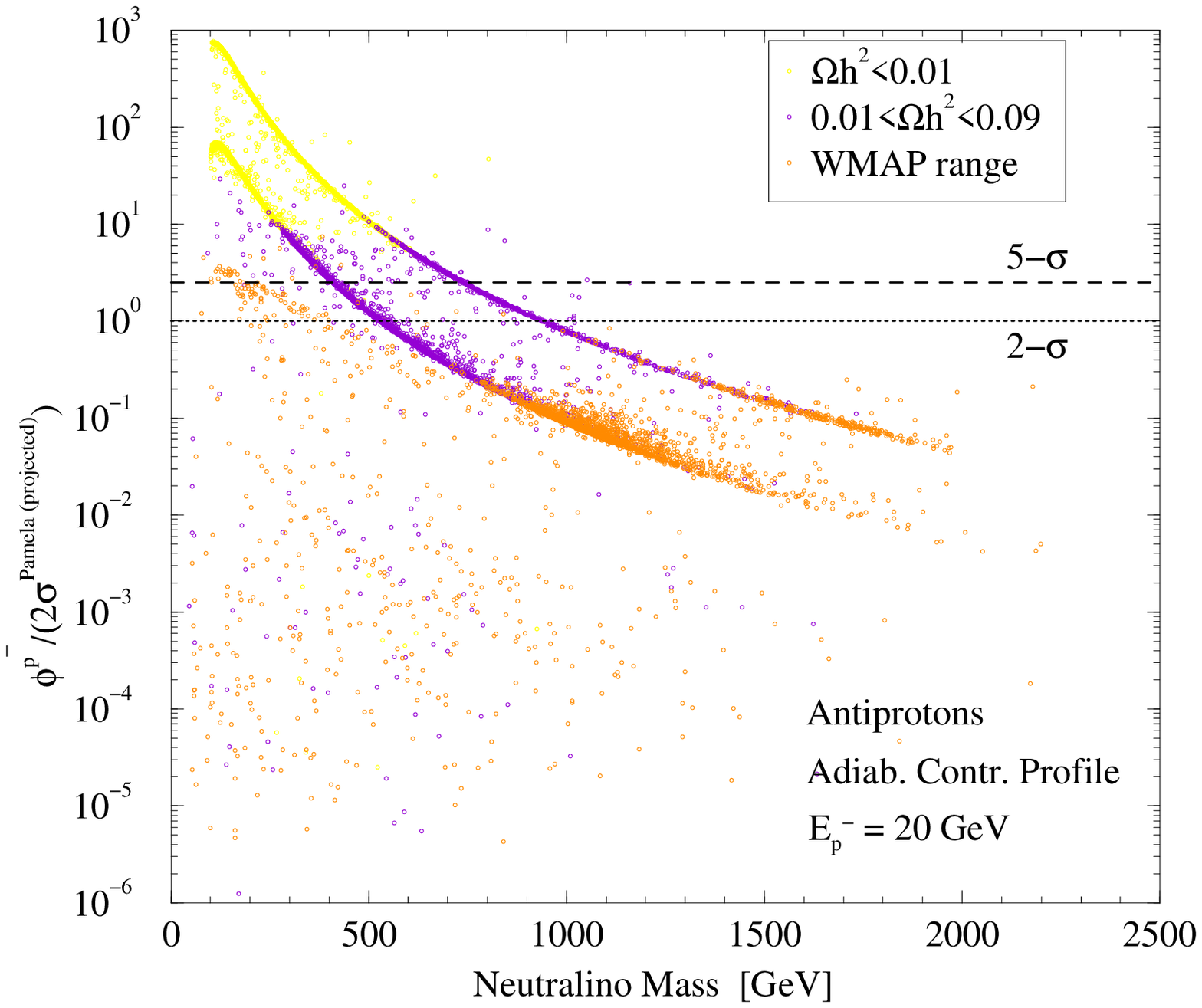}\\
({\em a\ }) & ({\em b\ })\\
\end{tabular}
\end{center}
\caption{\small \em Current and future discrimination sensitivities for antiproton fluxes, for a sample of models grouped by relic abundance.}\label{fig:PBARscat}
\end{figure*}
\begin{figure*}[!th]
\begin{center}
\includegraphics[scale=0.8]{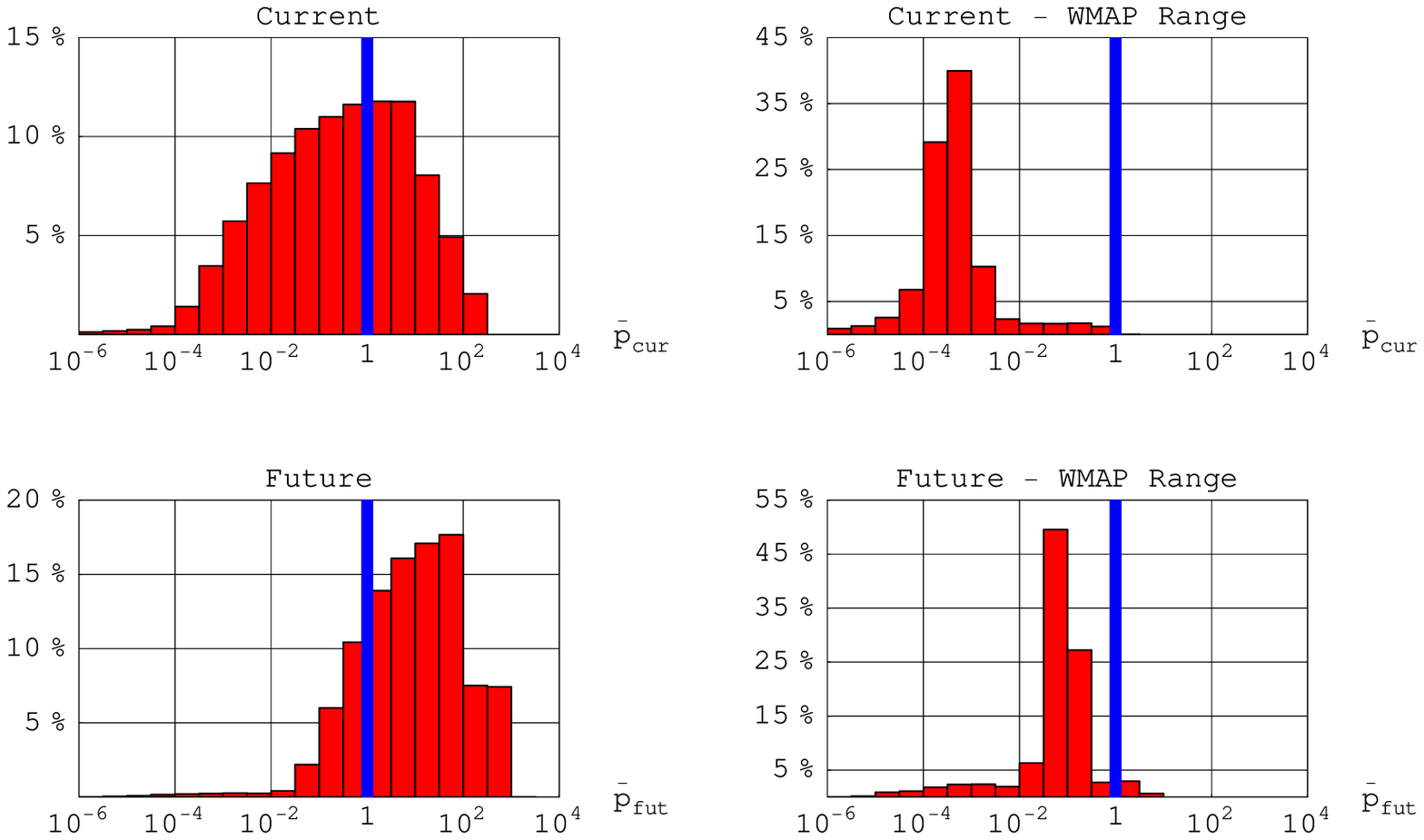}\\
\end{center}
\caption{\small \em A statistical analysis of the Visibility Ratio (primary flux over twice the error bars) for {\em antiprotons} with the {\em Adiabatically contracted} halo profile, for all models with a sufficiently low relic abundance  (plots to the left) and for models with a relic abundance in the WMAP range (plots to the right). The two upper plots refer to the current data at $E_{\overline p}=1.95 $ GeV, while the two lower plots to the future PAMELA projected sensitivity at $E_{\overline p}=20$ GeV.}\label{fig:PBARhistoB}
\end{figure*}
\begin{figure*}[!bh]
\begin{center}
\includegraphics[scale=0.8]{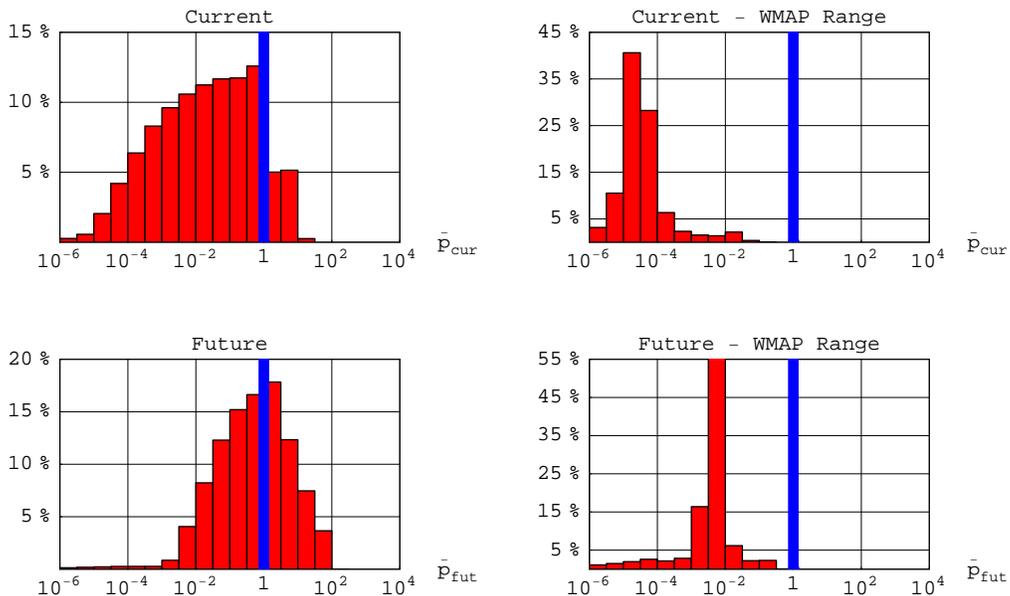}\\
\end{center}
\caption{\small \em The same statistical analysis for {\em antiprotons}, as in Fig.~\ref{fig:PBARhistoB}, but for the {\em Burkert profile}.}\label{fig:PBARhistoA}
\end{figure*}

The spectral features of primary antiprotons indicate a peak at energies of a few GeV, depending on the neutralino mass and composition. To this extent, we took the $E_{\bar p}=1.95$ GeV bin from the BESS-98 data \cite{bess}, which turns out also to be very well fitted by the computed secondary flux. As regards future perspectives, we took a putative $E_{\bar p}=20$ GeV bin at the PAMELA experiment, for which we simulated the experimental sensitivity and the data binning after three years of data taking, following Ref.~\cite{pamelasens}: our choice is in this case motivated by the fact that space based experiments will take data in the large kinetic energy range, where balloon-borne experiments could not be sensitive enough, a region where the signal-to-background ratio is particularly large. We limited ourselves, however, to 20 GeV since our scan in principle allows for neutralino masses as low as around twice that value. For positrons, we chose the $E_{e^+}=11.09$ GeV bin from the data of the HEAT-94/95 flights \cite{heat}, and took $E_{e^+}=23$ GeV for the PAMELA experiment. 

\begin{figure*}[!b]
\begin{center}
\begin{tabular}{cc}
\includegraphics[scale=0.5]{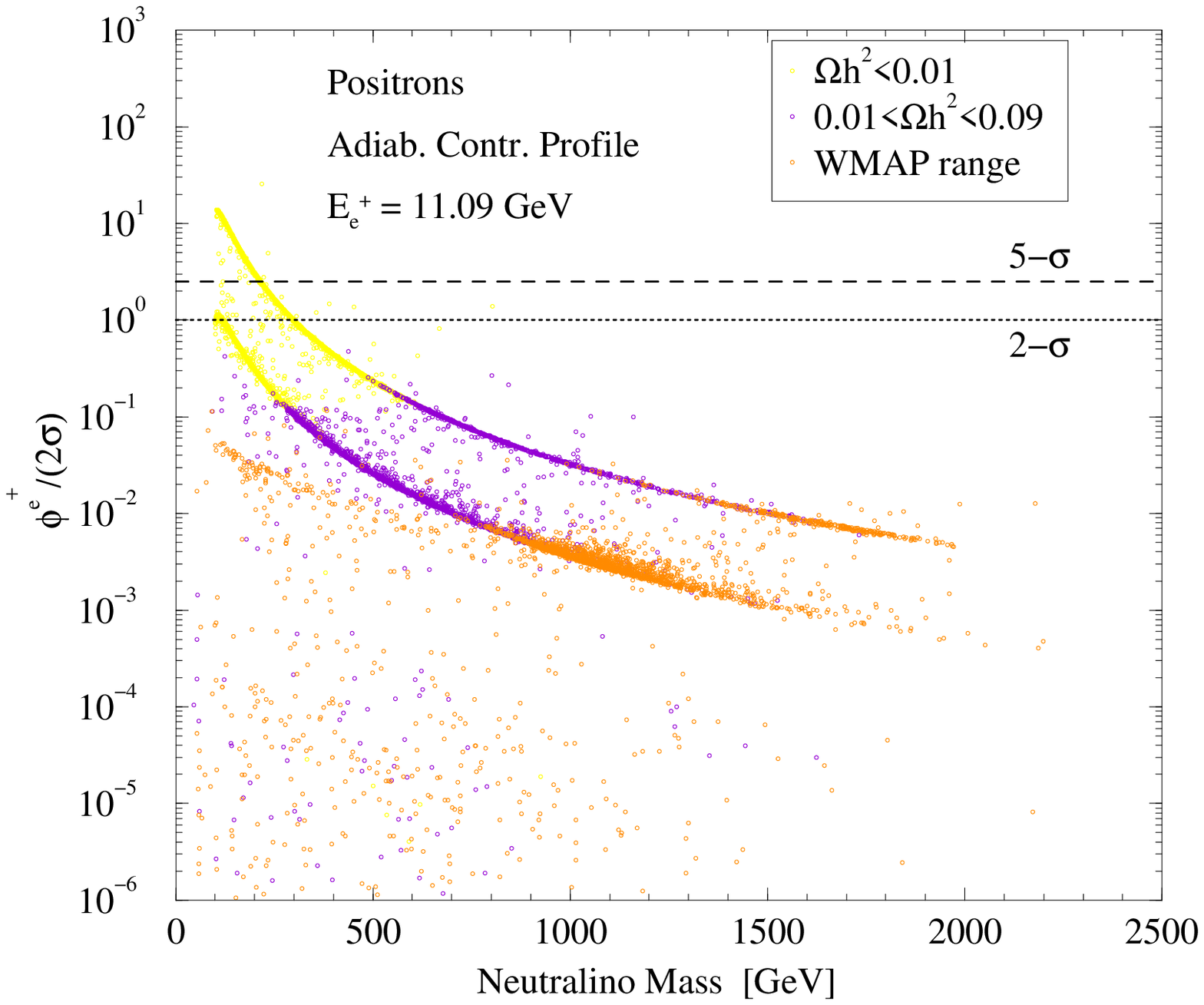} & \includegraphics[scale=0.5]{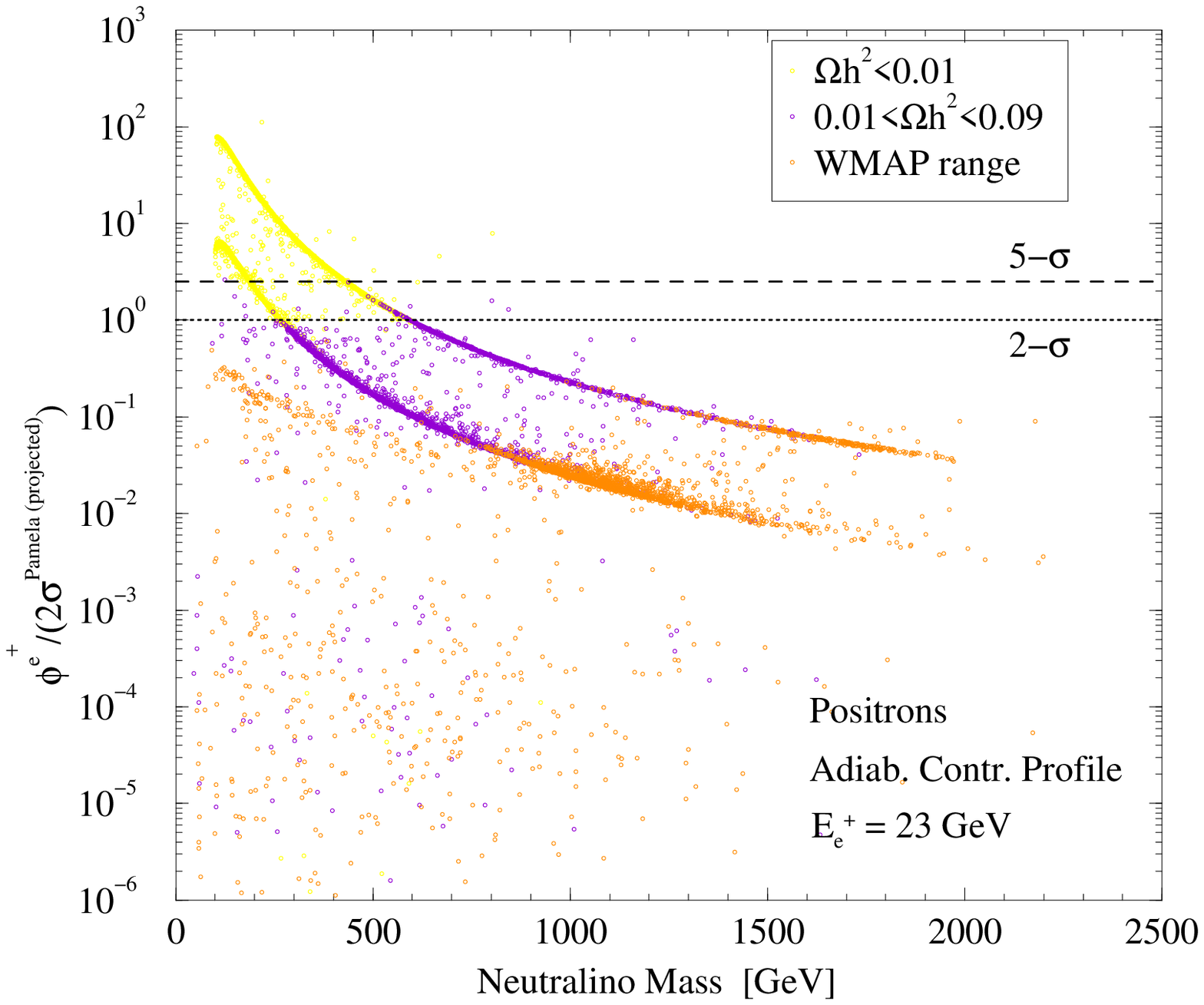}\\
({\em a\ }) & ({\em b\ })\\
\end{tabular}
\end{center}
\caption{\small \em Current and future discrimination sensitivities for positron primary fluxes, for a sample of models grouped by relic abundance.}\label{fig:EPLUSscat}
\end{figure*}
\begin{figure*}[!th]
\begin{center}
\includegraphics[scale=0.8]{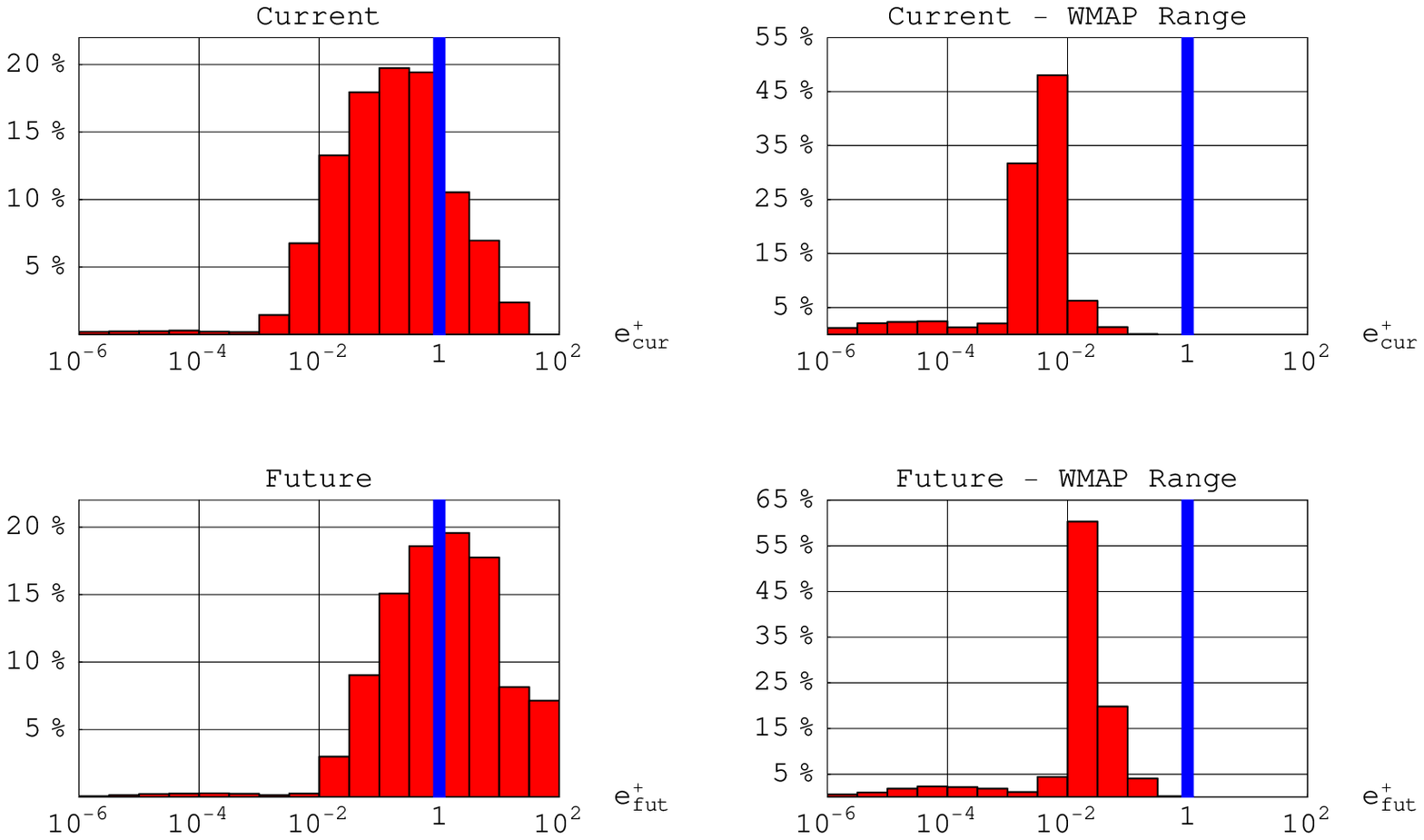}\\
\end{center}
\caption{\small \em A statistical analysis of the Visibility Ratio (primary flux over twice the error bars) for {\em positrons} with the {\em Adiabatically contracted} halo profile, for all models with a sufficiently low relic abundance  (plots to the left) and for models with a relic abundance in the WMAP range (plots to the right). The two upper plots refer to the current data at $E_{e^+}=11.09 $ GeV, while the two lower plots to the future PAMELA projected sensitivity at $E_{e^+}=23$ GeV.}\label{fig:EPLUShistoB}
\end{figure*}
\begin{figure*}[!bh]
\begin{center}
\includegraphics[scale=0.8]{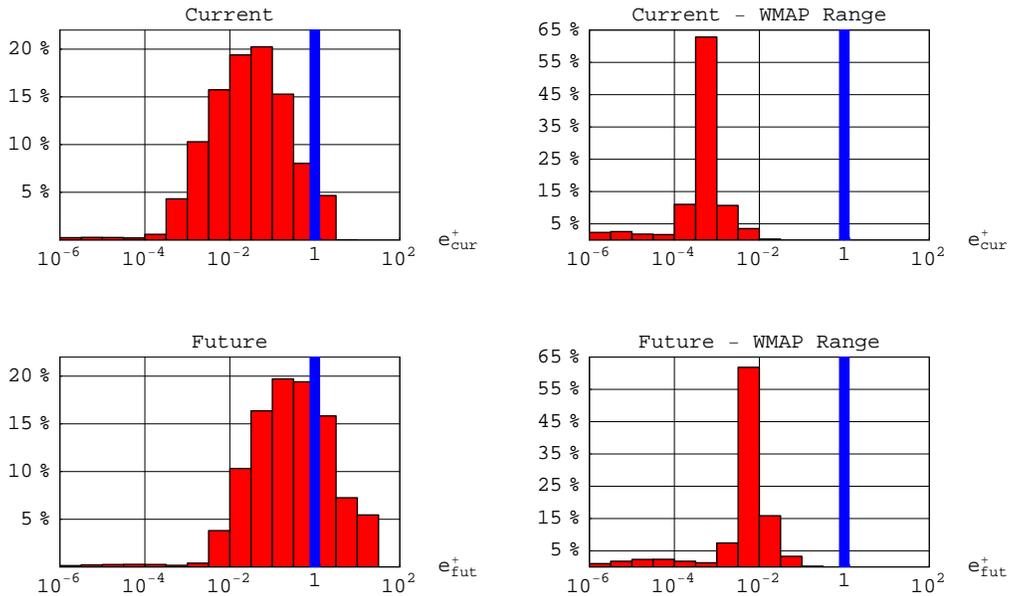}\\
\end{center}
\caption{\small \em The same statistical analysis for {\em positrons}, as in Fig.~\ref{fig:EPLUShistoB}, but for the {\em Burkert profile}.}\label{fig:EPLUShistoA}
\end{figure*}

Fig.~\ref{fig:PBARscat} collects our results concerning antiprotons, in terms of the $\phi^{\bar{p}_{\rm SUSY}}/(2\sigma_{\rm exp})$ ratio as a function of the neutralino mass, for current data (left) and at the forthcoming PAMELA experiment (right). We show here the results for the adiabatic halo profile. We also indicate the $5-\sigma$ line, corresponding to a SUSY signal exceeding five times the current experimental sensitivity. The color code indicates different neutralino thermal relic abundances as computed in the standard cosmological scenario; as pointed out for three particular benchmark scenarios in Ref.~\cite{Profumo:2004ty}, only models with large annihilation rates can give measurable antimatter signals. In fact, models with thermal relic abundances $\Omega_\chi h^2\lesssim 0.01$ are largely excluded by current data, and they will be thoroughly probed at future experiments. On the other hand, models within the WMAP range are not already constrained; only few of these models will be accessible in the future for neutralino masses up to 600-700 GeV. Notice the clear hierarchy of wino (upper line clustering) and higgsino-like (lower line clustering) neutralinos, which, due to the large branching ratio into gauge bosons final states, tend to have homogeneous antiproton fluxes, which only scale with the neutralino mass. This clearly allows model-independent predictions for these neutralino composition. Noticeably, a very large spread is instead present in the case of binos, where the dependence on the SUSY mass spectrum is critical for the computation of the antimatter flux.

Our statistical analysis, for the antiproton fluxes, is shown in Fig.~\ref{fig:PBARhistoB} and Fig.~\ref{fig:PBARhistoA} respectively for the Burkert and for the adiabatically contracted profile. With both halo profiles, models with large annihilation rates will be largely accessible to future space based experiments. On the other hand, models with thermal relic abundances within the WMAP range are not currently constrained, and will be one to two orders of magnitude below future sensitivity. We therefore stress that {\em supersymmetric models yielding the correct dark matter thermal relic abundance in the standard cosmological scenario do not produce antimatter fluxes detectable at future space based experiments}.

\begin{figure*}[!b]
\begin{center}
\begin{tabular}{cc}
\includegraphics[scale=0.5]{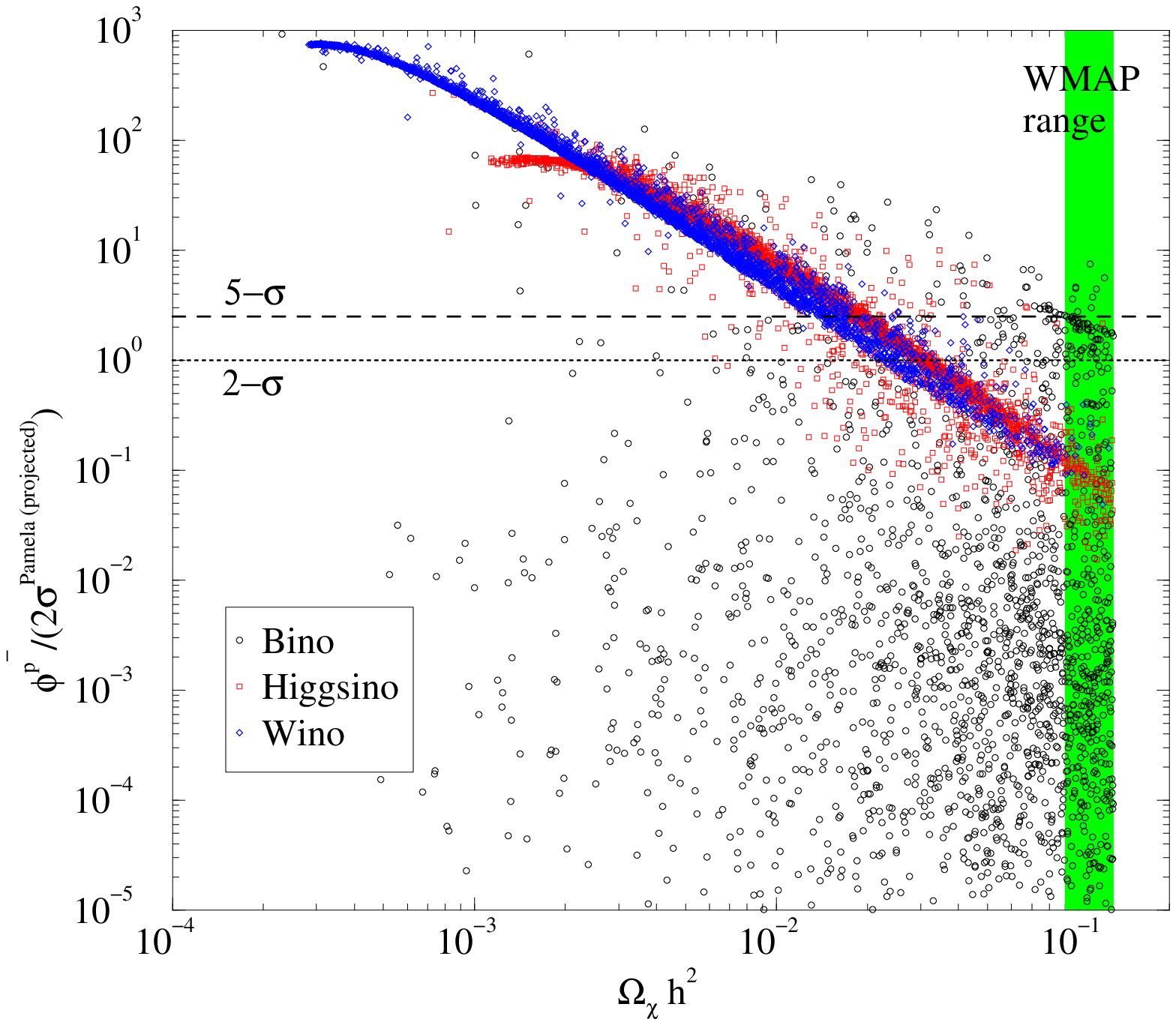} & \includegraphics[scale=0.5]{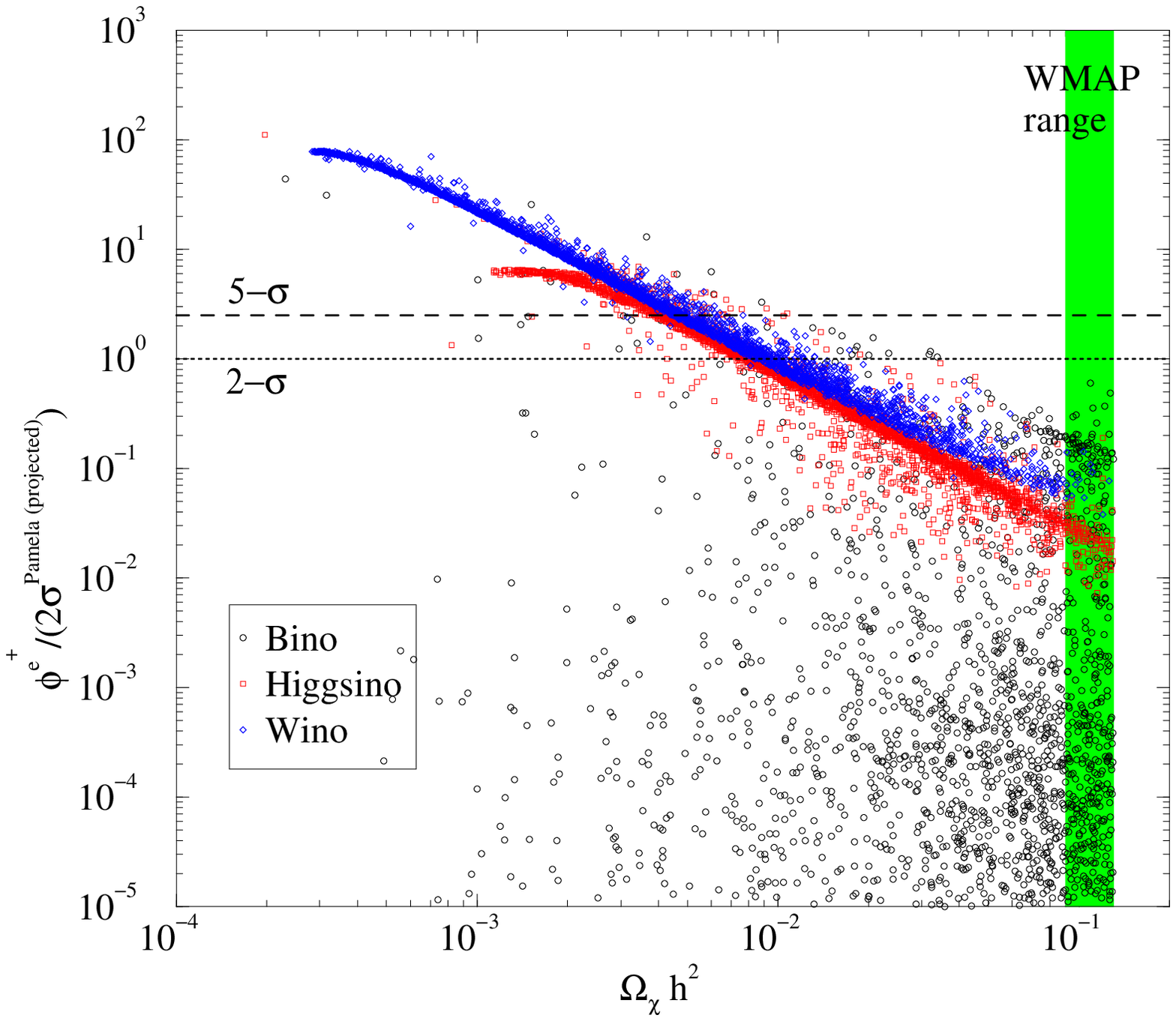}\\
({\em a\ }) & ({\em b\ })\\
\end{tabular}
\end{center}
\caption{\small \em The correlation between the neutralino relic abundance and the projected sensitivities on antiproton (a) and positron (b) fluxes. Models are grouped according to the nature of the lightest neutralino. The WMAP range is indicated by a vertical green strip.}\label{fig:CORR_O}
\end{figure*}

\begin{figure*}[!t]
\begin{center}
\includegraphics[scale=0.65]{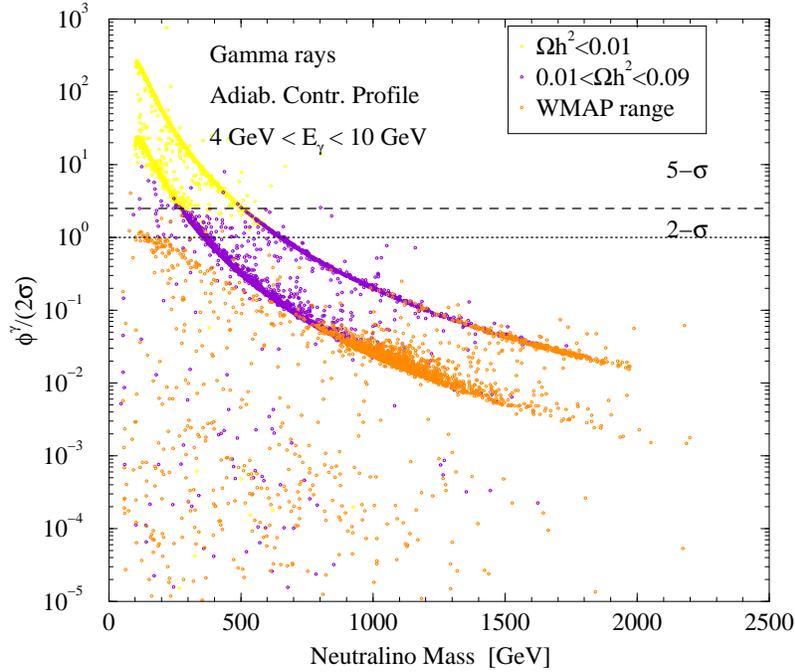}\\
\end{center}
\caption{\small \em A scatter plot for the ratio between the expected gamma ray flux from neutralino annihilations in the galactic center and the 2-$\sigma$ sensitivity in the bin at the largest energies sampled by EGRET. Models are grouped according to their relic abundance.}\label{fig:SCATgammas}
\end{figure*}
\begin{figure*}[!t]
\begin{center}
\includegraphics[scale=0.85]{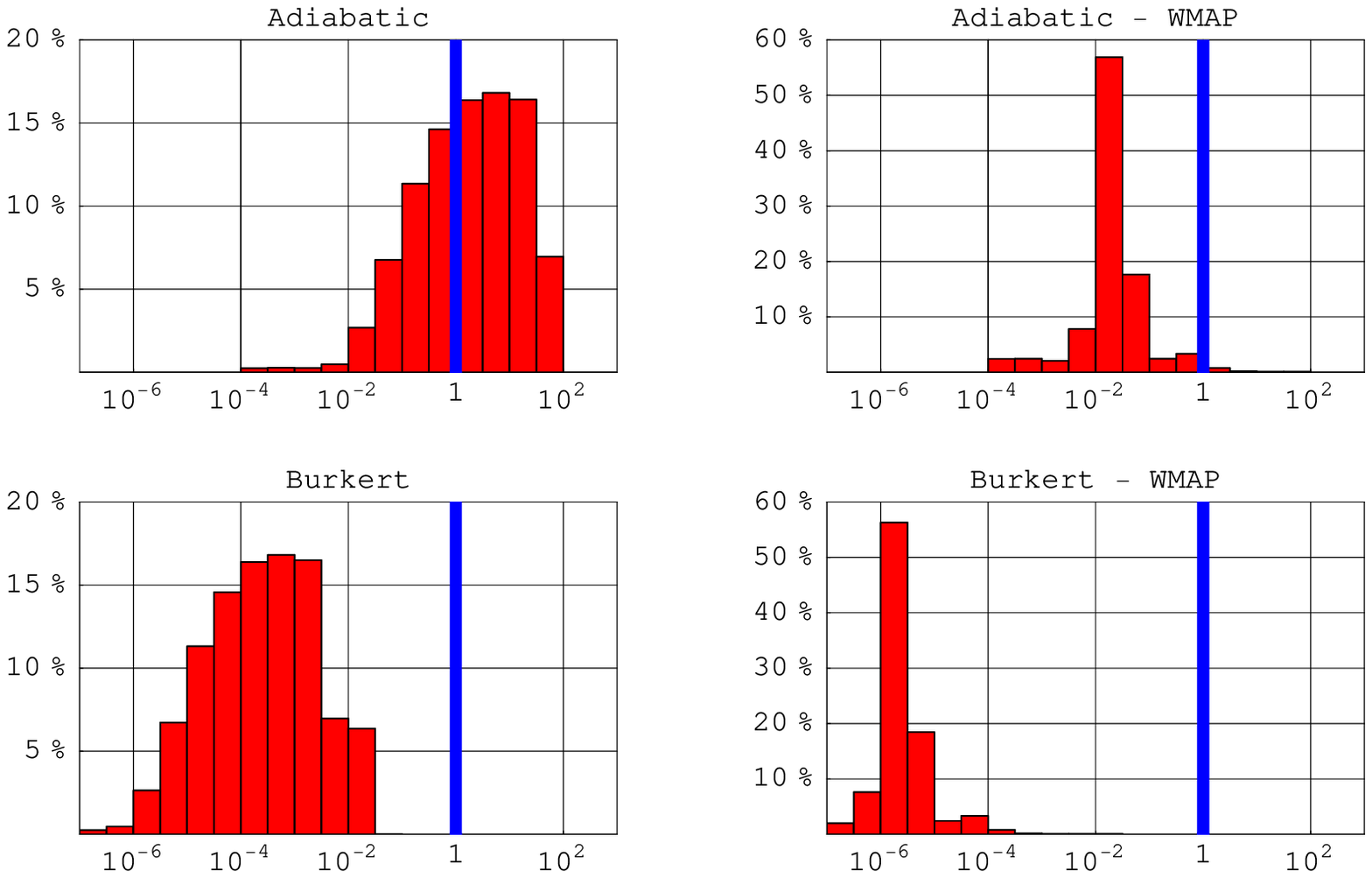}\\
\end{center}
\caption{\small \em A statistical analysis of the Visibility Ratio (flux over twice the error bars) for {\em gamma rays}, for all models with a sufficiently low relic abundance  (plots to the left) and for models with a relic abundance in the WMAP range (plots to the right). The two upper plots refer to the Adiabatically contracted profile, while the two lower plots to the Burkert profile.}\label{fig:HISTOgammas}
\end{figure*}

The situation for positrons is rather similar to that of antiprotons, though, less promising. The scatter plots of Fig.~\ref{fig:EPLUSscat} indicate that we will be able to detect, at best, models featuring a thermal relic abundance up to $\Omega_\chi h^2<0.01$. The discrimination ratios for models within the WMAP range fall, in the best case scenario, approximately three orders of magnitude below future sensitivities. The histograms of Fig.~\ref{fig:EPLUShistoB} and \ref{fig:EPLUShistoA} summarize the statistics we collected for the positron channel.

To further detail on the correlation between thermal relic abundance and antimatter fluxes, and to compare antiproton and positron fluxes, we plot in Fig.~\ref{fig:CORR_O} a sample of models, grouped by neutralino composition, in the $(\Omega_\chi h^2,\phi^{\bar a}/(2\sigma))$ plane. The case of antiprotons is showed in the left panel, while that of positrons in the right panel. First, notice that at a given $\Omega_\chi h^2$, the discrimination ratio for antiprotons is everywhere one order of magnitude better than that of positrons. Secondly, a clear correlation between $\Omega_\chi h^2$ and $\phi^{\bar a}/(2\sigma)$ is present: Wino and higgsino like neutralinos lie on a relatively thin strip in this plane, which ends, in the WMAP preferred range, more than one order of magnitude below the visibility lines both for positrons and antiprotons detection. Binos scatter instead over a broad range of fluxes. In this respect, we generalize here the conclusions of Ref.~\cite{Profumo:2004ty}, in that antimatter searches are a promising road towards indirect discovery of neutralinos provided some relic density enhancement mechanism is operative, and models with a large neutralino annihilation rate may account for the whole WMAP-inferred amount of dark matter.

In the presence of a halo profile featuring a cusp in the Galactic Center (GC), a sizeable contribution to the $\gamma$-ray flux is expected to be provided by neutralino annihilations. The EGRET experiment, on the Compton Gamma-Ray Observatory, has reported a tentatively extended $\gamma$-ray source $\sim 1.5^{\circ}$ from the GC \cite{MH}, with a spectrum incompatible with the one expected from interactions of primary cosmic rays with the interstellar medium, and compatible, on the other hand, with a WIMP-induced component \cite{cesarini}. Though alternative explanations for this so-called {\em EGRET excess} have been put forward, these measurements may be turned around and used as a constraint, for a given halo profile, on the SUSY $\gamma$-ray contribution, and thus on SUSY models \cite{Profumo:2004ty}. The most stringent bound comes from the largest energy bin, with photon energies in the range $4\, {\rm GeV}<E_\gamma<10\, {\rm GeV}$, where the standard background component is largely suppressed with respect to the measured $\gamma$-ray flux. In this respect, we directly compare the SUSY contribution with the $2-\sigma$ bound from the EGRET measurement, and therefore declare not viable (within a given halo profile) any SUSY model yielding a flux larger than the sum of the central value and twice the error bars. Our results, for the  adiabatically contracted halo profile, are collected in Fig.~\ref{fig:SCATgammas}. Once again we notice the usual $\sim m_\chi^{-2}$ suppression, the clustering of the fluxes for wino and higgsino like neutralinos, and the correlation with the thermal relic abundance. Models within the WMAP range are not constrained by current data, while low relic density models with $\Omega_\chi h^2<0.01$ are excluded. The statistical analysis reported in Fig.~\ref{fig:HISTOgammas} highlights the large dependence of $\gamma$-ray results on the halo profile: assuming a cored profile (two lower histograms) this indirect detection channel fails to give any constraint on the SUSY parameter space, even in the SUSY models with the largest annihilation rate. Moreover (see the two histograms to the right) models in the WMAP range are quite far from being detectable, even in the most optimistic halo profile setup. Future measurements of the mentioned $\gamma$-ray source may either put more stringent constraint, or confirm alternative explanations for the EGRET excess \cite{pohl}.

\subsection{An Overview of SUSY dark matter Search Strategies}

In view of the previous considerations, we quantitatively summarize in Tab.~\ref{tab:overiew} our results as regards the comparison of different supersymmetric dark matter detection strategies. We made use of two different, benchmark halo models, a cuspy profile ({\em Adiabatically contracted profile}) and a cored profile ({\em Burkert profile}), as outlined in \cite{pierohalos}. Results concerning direct detection and neutrino telescopes are very mildly affected by the halo model under consideration, and we therefore reported our results only for the Adiabatically contracted profile.

In the column {\em All Models} we include both models whose thermal relic abundance falls in the WMAP preferred range and those with lower relic densities: we recall that {\em we do not perform any neutralino density rescaling}, under the hypothesis that some relic density enhancement or non-thermal production mechanism allows models with large annihilation rates to be compatible with the correct required amount of dark matter.

{\large
\begin{table}[!t]
\begin{center}
\begin{tabular}{c|c|c|c|c|}
 & \multicolumn{2}{c}{\bf All Models} &  \multicolumn{2}{c}{\bf WMAP Range}\\
\hline
&&&&\\
 & {\bf Current} & {\bf Future}  & {\bf Current} & {\bf Future} \\
&&&&\\
\hline
&&&&\\
Direct SI & $\sim 0$\% & 22\% &  $\sim 0$\% & 20\% \\
&&&&\\
$\mu$ f. Sun & $\sim 0$\% & 5\% &  $\sim 0$\% & 3\% \\
&&&&\\
\hline
&&&&\\
$\overline{p}$ -- Adiab.~P. & 38\% & 80\% &  $\sim 0$\% & 4\% \\
&&&&\\
$e^+$ -- Adiab.~P. & 20\% & 52\% &  $\sim 0$\% & $\sim 0$\% \\
&&&&\\
\hline
&&&&\\
$\overline{p}$ -- Burkert~P. & 10\% & 41\% &  $\sim 0$\% & $\sim 0$\% \\
&&&&\\
$e^+$ -- Burkert~P. & 5\% & 28\% &  $\sim 0$\% & $\sim 0$\% \\
&&&&\\
\hline
\end{tabular}
\end{center}
\caption{\small \em A summary of the statistical analysis of the various dark matter detection methods. The first two columns refer to all models (WMAP+low relic density), while the last two columns to models with relic abundance within the WMAP range only. For antimatter and gamma rays we indicate the results for both the Adiabatically contracted profile and for the Burkert profile.}\label{tab:overiew}
\end{table}
}

Let us stress, first of all, that, in a statistical sense, {\em the bulk of the supersymmetric parameter space compatible with the lower and upper WMAP bounds on the neutralino relic abundance has still not been probed by dark matter searches}, as highlighted by the zeroes appearing in the third column. On the other hand, {\em taking into account low relic density models, only antimatter searches are currently providing significant constraints on SUSY models}, though with a large dependence on the assumed halo profile. On the other hand, neutrino telescopes and direct detection  rule out only marginal portions of parameter space\footnote{This does not mean, of course, that there are no parameter space choices whose detection rates at direct searches and neutrino telescopes fall above current exclusion limits: what we find is that these choices are statistically marginal.}.

As regards future prospects, we find that a significant portion of the viable SUSY parameter space will be probed at large direct detection facilities (approximately $20\%$, quite independently of the relic abundance of the models). Perspectives at Neutrino Telescopes are less exciting, but we find that still around 5\% of the parameter space will be accessible at IceCube. Antimatter searches will fail to provide any strong constraint on models with thermal relic abundance in the WMAP range; however, provided some relic density enhancement mechanism is operative, they could become a prominent road to SUSY dark matter discovery on future space based experiments (AMS, PAMELA). Antiprotons are found to be in any case a more promising dark matter indirect detection channel than positrons.

\section{Conclusions}\label{sec:concl}

The present paper is devoted to a statistical analysis of the general minimal, flavor diagonal, supersymmetric extension of the standard model. Resorting to a random scan of the MSSM parameter space, in the soft breaking mass range between 50 GeV and 5 TeV, we first studied the relative effectiveness of the various phenomenological constraints. We singled out $10^5$ {\em viable models}, also consistent with the upper WMAP bound, and performed a statistical study out of them, starting from the expected values of \bsg, $m_h$ and $(g-2)_\mu$. We found that the $CP$ even lightest Higgs boson mass is limited from above at about 132 GeV. The SUSY contributions to the muon anomalous magnetic moment have been shown to be always noticeably suppressed.

We then turned to the topic of supersymmetric dark matter. We showed that the relic abundance of the viable models tend to be around one order of magnitude below the WMAP preferred range. This is due to the large number of models featuring a wino or a higgsino like lightest neutralino. If one requires the thermal relic abundance to be {\em within} the WMAP $2-\sigma$ range, one finds that pure higgsino states masses cluster around 1 TeV, while winos around 1.6 TeV, killing any hope of detecting supersymmetric signals at the LHC. On the other hand, bino like neutralinos with the correct relic abundance feature, in the large majority of cases, a coannihilating partner, or a resonant annihilation cross section, and the resulting LSP mass spreads over a wide range. We analyzed the nature of coannihilating partners and the relative efficiency in the reduction of the bino relic abundance, and provided details on resonant bino annihilations into heavy Higgs bosons.

Is it therefore correct to consider minimal supergravity as a benchmark MSSM scenario? The answer is probably not so straightforward. We found that models whose relic abundance lies in the WMAP range have in most instances a higgsino like neutralino, but that many models also feature a bino or a wino LSP: in this respect, mSUGRA introduces some kind of bias in predicting, over most of its parameter space, a bino-like neutralino. As in mSUGRA, binos in the MSSM require the existence of relic density suppression mechanisms, like coannihilations or resonances: however, the general MSSM allows for a much richer plethora of coannihilating partner, and, consequently, a wider neutralino mass range.

The second part of our analysis has been devoted to a comparison of supersymmetric detection strategies. We made use of two benchmark halo models, in which the density and velocity distributions have been self-consistently computed. The analysis has been carried out with the use of {\em Visibility Ratios}, which allowed an assessment of current and future detection channels and a comparison of different search strategies. 

We found that current data on dark matter searches provide weak constraints on the bulk of the general MSSM viable models. Future perspectives, for models whose thermal relic abundance lies within the WMAP range, appear to be quite promising for direct, spin-independent searches (we find that $\sim$20\% of viable models will be within future sensitivity) and, to a less extent, for neutrino telescopes looking at neutrinos produce by neutralino annihilations in the Sun. Antimatter searches were found to be largely dependent on the neutralino relic abundance: models within the WMAP range do not typically produce antimatter fluxes which will be detectable at future space based experiments. Taking into account the possibility of relic density enhancement mechanisms in the Early Universe, models with large annihilation rates could account for the whole of dark matter: in this case, antimatter searches put severe constraints on most SUSY models, and may become soon a prominent road towards dark matter discovery. This conclusion holds true quite independently of the details of the dark matter distribution in the inner part of our Galaxy, contrary to the case of gamma rays searches. Finally, we provided a summary table in which we quantify the statistical relevance of a given dark matter search strategy, both for low relic density models and for models with WMAP compatible neutralino thermal abundance.

\section*{Acknowledgements} It is a pleasure to acknowledge Piero Ullio for useful discussions and suggestions.


%

\end{document}